# Human-AI Synergy in Adaptive Active Learning for Continuous Lithium Carbonate Crystallization Optimization

Shayan Mousavi Masouleh [1, 2], Corey A. Sanz [3], Ryan P. Jansonius [3], Cara Cronin [3], Jason E. Hein [3], Jason Hattrick-Simpers [1, *]

[1] Canmet MATERIALS, Natural Resources Canada, 183 Longwood Rd S, Hamilton, ON, Canada

[2] Clean Energy Innovation, National Research Council of Canada, 2620 Speakman Dr, Mississauga, ON, Canada

[3] Telescope Innovations, 301-2386 E Mall, Vancouver, BC, Canada

**Corresponding Author**

* jason.hattrick-simpers@nrcan-rncan.gc.ca

**ABSTRACT**

As the demand for high-purity lithium surges, primarily fueled by the adaptation of the electric vehicle (EV) industry, the need for cost-effective extraction and purification technologies intensifies. The Smackover Formation in southern Arkansas, recently identified as one of the world's largest lithium resources, offers vast potential. This formation is part of a broader array of lithium resources across North America, many of which possess lower-grade lithium compared to renowned sources like South American brines. These alternative formations, while presenting significant opportunities, require innovative purification techniques to make their exploitation economically viable. Continuous crystallization is a promising method to produce battery-grade lithium carbonate from these lower-grade sources. Yet, the optimization of this process is challenging due to its complex parameter space, often constrained by scarce data. This study introduces a Human-in-the-Loop (HITL) assisted active learning framework aimed at adapting and optimizing the continuous crystallization process of lithium carbonate. By integrating human expertise with data-driven insights, this approach significantly accelerates the optimization of lithium extraction from challenging sources. Our results demonstrate the framework's ability to rapidly adapt to new data, improving the process's tolerance to critical impurities, such as magnesium, by industry practices at a few hundred ppm, and extending it to handle contamination levels as high as 6000 ppm. This makes the use of low-grade lithium resources contaminated with such impurities feasible, potentially reducing overhead processes. By leveraging artificial intelligence, we not only refined the operational parameters but also demonstrated a potentially reduced need for extensive pre-refinement, promoting the use of lower-grade materials without sacrificing product quality. This advancement marks a significant step towards economically harnessing North America's lithium reserves, particularly those in the Smackover Formation, thereby contributing to the sustainability of the lithium supply.

**TOC GRAPHICS**

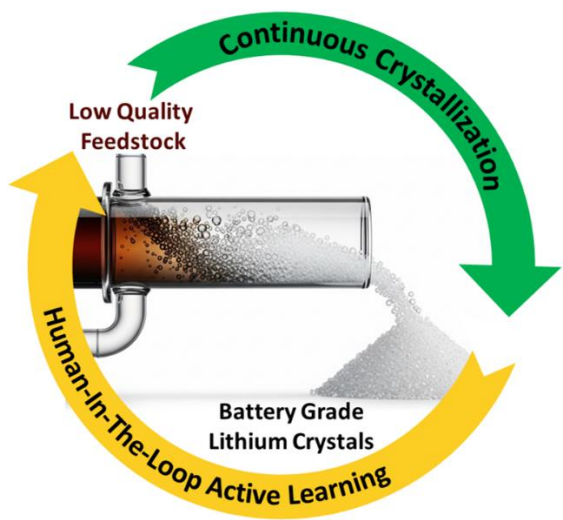

# 1. Introduction

Demand for high-purity lithium (Li), essential for batteries in electronics and vehicles, is projected to outstrip supply by 2035 [1–3]. As the demand continues to escalate, lithium-ion battery demand is expected to grow by 27% annually [4]. This volatility underscores the urgency for more economically viable extraction methods as the traditional techniques remain cost-intensive and environmentally taxing [5]. While total global lithium resources are estimated at 527 million tonnes—far exceeding the annual consumption rate of 0.5 million tonnes—much of this lithium is sourced from low-grade deposits that are expensive to exploit due to their complex geochemistry and the presence of impurities [1,2].

Among these low-grade sources, the Smackover Formation in southern Arkansas stands out as a significant resource. Identified as potentially one of the world's largest lithium reserves, the Smackover Formation features high concentrations of lithium in brines—over 400 milligrams per liter—which are currently brought to the surface as waste streams from the oil, gas, and bromine industries [6–10].

However, exploiting these resources is not straightforward. The Smackover's lithium-bearing brines are characterized by a high ratio of impurities to lithium, approximately 1000 atoms of impurity for every atom of lithium. These impurities include elements such as sodium (Na), potassium (K), magnesium (Mg), and calcium (Ca), closely related to lithium on the periodic table, which complicates their separation due to their similar solubility, charge, and mass properties [6]. This challenge necessitates a shift from traditional methods to innovative extraction techniques that are both economically viable and environmentally sustainable.

Traditional methods for producing battery-grade $Li_2CO_3$ are costly and involve multiple steps, including lengthy evaporation processes with high water usage [11–14]. Exploiting lower-grade brines would be even more expensive due to their dilution, necessitating process improvements to reduce reagent, solvent, and water usage [2,15]. In contrast, continuous crystallization—a technique commonly used in the pharmaceutical industry but rarely in mining—is well-suited for this purpose [16,17]. This method achieves metal salt purities of 90-99.9% and eliminates the need for evaporative concentration [18–24], thus, reducing water and land use as well as production time [15]. According to a United States Department of Energy assessment, direct lithium extraction using continuous crystallization can cut production costs by 24% compared to traditional evaporative methods, enabling cost-effective purification of $Li_2CO_3$ from low-grade brines in a single step [2,15].

Nonetheless, the continuous crystallization process is not without its challenges, particularly when applied to the low-grade lithium brines such as those found in the Smackover Formation [6,7,19]. In this regard, the primary challenge is not merely concentrating the lithium, which can be achieved through methods like evaporation, reverse osmosis, or solar concentration, but effectively rejecting closely related impurities (Na, K, Ca, Mg). To achieve high-purity lithium carbonate ($Li_2CO_3$), the crystallization process must be finely tuned: a dilute solution might lead to sparse crystal formation and low mass recovery, while a too concentrated solution can cause rapid nucleation, trapping impurities within the crystals. Adding to the complexity, $Li_2CO_3$ exhibits inverse solubility, decreasing from 13 g/L at 20 °C to 8.6 g/L at 80 °C. Thus, continuous crystallization needs to be adapted to be either impurity-tolerant or highly selective for lithium, minimizing the need to remove every challenging impurity. Optimizing such intricate chemical operations is a formidable task due to the high dimensionality of the control space and the sophisticated chemistry involved.

Traditional optimization strategies, such as conventional design of experiment (DOE) methodologies, often involve exploring a vast experimental space. In our study, we initially identified 10 critical variables, which, under a full factorial DOE, would necessitate about more than approximately 1,024 experiments. Given our experimental throughput constraints, limited to about four per week, conducting such a large number of experiments was impractical. This limitation highlighted the need for an alternative approach that could achieve optimization with fewer experiments.

Over the past decade, various AI-driven active learning solutions have emerged, treating the optimization task as a search through a "black box." These methods have led to more efficient exploration and optimization of complex systems [25–31]. Nonetheless, in systems with a moderate to high number of dimensions, these methods can struggle to swiftly optimize the systems. Typically, AI-driven active learning approaches rely heavily on machine learning models, which build correlations without necessarily providing causal understanding or integrating well-established, yet hard-to-quantify, heuristics of the physical and chemical nature of the system under study. This reliance can result in a process that remains time-consuming and resource-intensive. Furthermore, design biases may arise from a limited understanding of the system's complexities, leading to skewed outcomes and compounding errors. These biases can hinder the model's ability to generalize, impacting its overall performance and reliability.

To mitigate biases and address complexity challenges, the field of human-in-the-loop AI has emerged as a promising solution [32–34]. This approach leverages the collaboration between human intelligence and artificial intelligence [32]. Human cognitive abilities and domain expertise play a crucial role in enhancing AI's predictive capabilities by interpreting data-driven correlations and offering intuition-driven insights. These insights help in refining the evaluation process of AI models, ensuring that the models align more closely with real-world complexities and expectations. Human-in-the-loop AI involves human input in tasks such as data collection, algorithm selection, and model tuning, creating a feedback cycle that helps reduce biases in both human decision-making and AI predictions [33]. This collaborative framework allows for quick adjustments and a deeper understanding of the machine learning workflow, ensuring that AI-driven systems become more adaptive, efficient, and effective in managing complex optimization tasks. This integration not only addresses the limitations of each approach but also combines their strengths to improve overall outcomes in high-dimensional optimization environments [32–34].

In this paper, we present a human-in-the-loop-assisted active learning (HITL-AL) AI framework specifically designed to optimize a continuous crystallization technique for producing high-quality, battery-grade lithium from diverse low-grade brines. This approach directly addresses the challenges posed by feedstocks like the Smackover Formation's complex geochemistry, where high impurity levels closely related to lithium complicate traditional extraction methods. Our objective extends beyond merely adjusting control parameters such as reactor temperatures and flow rates. It also involves investigating the interactions between brine compositions and system controls, particularly the interplay of major contaminants like Na, Ca, Mg, and K, to determine how much contamination and dilution can be tolerated while still producing battery-grade outcomes. By doing so, we aim to unlock the potential of these abundant but challenging resources, reducing reliance on high-quality sources and advancing cost-effective, sustainable lithium extraction methods.

In our HITL-AL process, human experts play a central role in refining machine learning-suggested experiments, using their judgment to focus on those most likely to yield meaningful results. This strategic selection is crucial for conducting experiments within practical throughput constraints while exploring promising pathways that models might overlook. Moreover, human experts are integral to evaluating outcomes and adjusting both hypotheses and workflows. This involves developing new hypotheses from emerging data and rigorously testing these ideas, helping to identify and correct biases in design and chemical assumptions, such as the difficulty of impurity removal and the ranges of control parameters. The methods section provides detailed insights into these processes and how different hypotheses are tested.

This iterative approach not only reduces the number of experiments required but also uncovers critical insights that drive innovation. For instance, through this collaborative and adaptive process, we discovered that adjusting cold reactor temperatures significantly reduces magnesium impurities. This counterintuitive breakthrough was achieved with minimal observations, demonstrating the effective synergy of human intuition and AI analysis. This finding significantly enhances the production of battery-grade lithium by expanding the acceptable range of magnesium contamination levels. We explore these discoveries further, illustrating the framework's effectiveness in optimizing lithium production. Additionally, we highlight the broader implications of our findings, showing how the integration of human expertise and AI not only improves experimental efficiency but also provides a sustainable and cost-effective solution to challenges posed by low concentration and polluted brines, such as those found in the Smackover Formation.

## 2. Methods

This study introduces a Human-in-the-Loop Active Learning (HITL-AL) framework designed to optimize a continuous lithium crystallization process from low-grade brines, specifically brines with high levels of sodium (Na), magnesium (Mg), calcium (Ca), and potassium (K). These impurities, prevalent in the Smackover Formation's brine sources, significantly complicate the extraction and purification of lithium. The primary objective is to ascertain operational conditions within our experimental setup that can efficiently convert these challenging brines into battery-grade lithium carbonate, even when heavily contaminated with these impurities (see **Table S1** for specifications of battery-grade lithium carbonate used in this study). Given our limited capacity to conduct experiments—restricted to approximately four per week due to resource and time constraints—it is crucial to maximize the efficiency and impact of each experimental run. Our integrated workflow synergizes experimental and computational efforts, facilitated by AI and human expertise, in a cyclic process that progressively refines our understanding and control of the crystallization system; see **Figure 1**. Detailed descriptions of the experimental setup are provided in section 2.1.

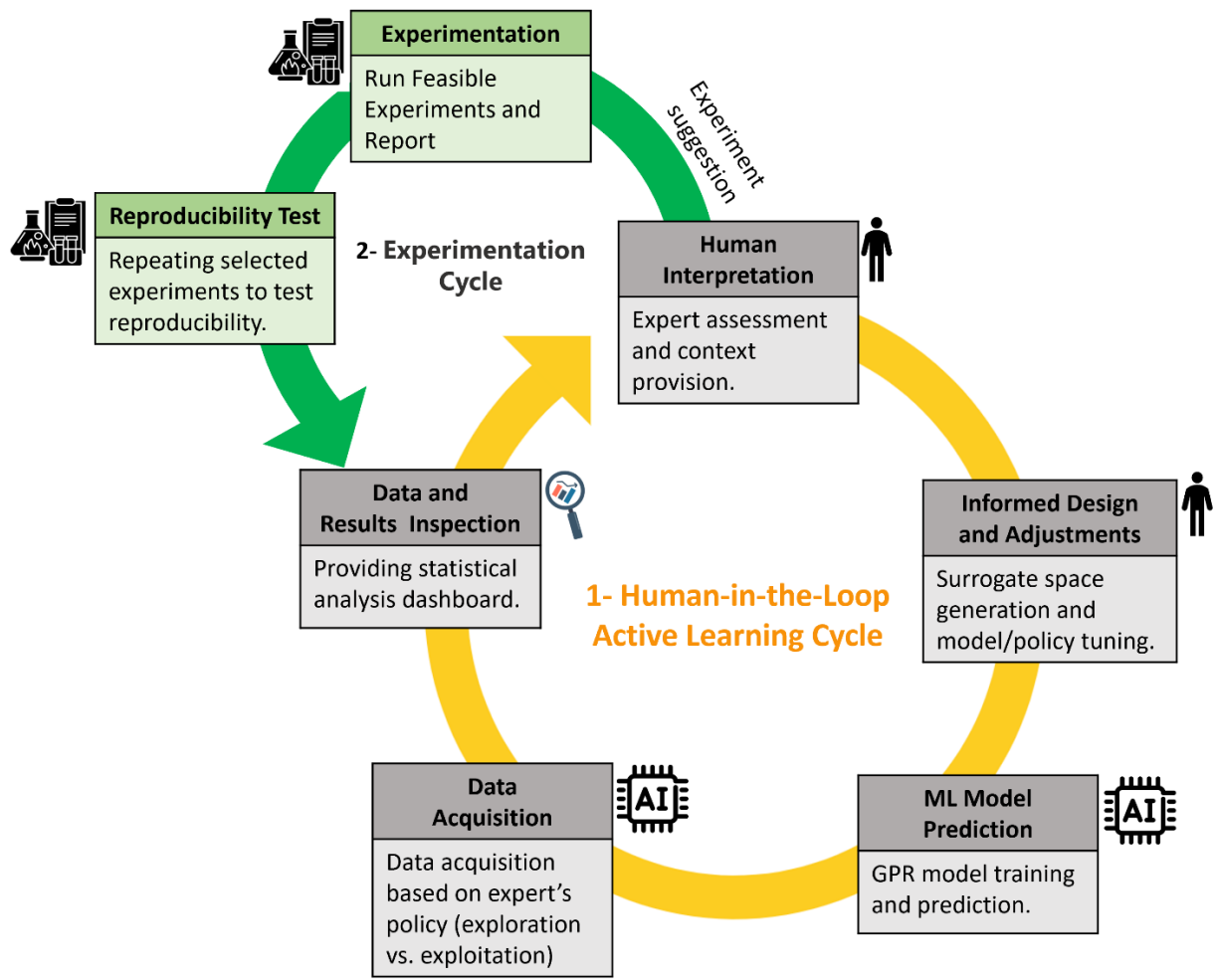

**Figure 1.** Overview of the integration between the experimentation cycle and the human-in-the-loop active learning cycle, highlighting the different steps or modules within each phase.

Upon initiating the HITL-AL cycle, as illustrated in **Figure 1**, the "Data and Results Inspection Module" acts as the initial juncture. This module integrates cumulative experimental observations into our workflow. It features a statistical dashboard that compiles and analyzes incoming data. Detailed information on these statistical analyses can be found in section 2.2.1. This dashboard provides human experts with insights into parameter interdependencies and correlations, crucial for real-time monitoring and timely response to experimental findings.

As outlined in **Figure 1**, following the inspection step, the "Human Interpretation" phase occurred. In this phase, experts apply their scientific and intuitive insights to the results of the statistical analysis. This step emphasizes the integration of human expertise into the cycle, thus avoiding the exhaustive need to fully automate complex decision-making in R&D setups.

Leading directly from the interpretation step, is the "Informed Design and Adjustment" phase. In this phase, the insights obtained from human interpretation are employed to meticulously adjust and define the configurations, ranges, and constraints within the surrogate space—a multidimensional representation of experimental conditions. Each point within this space specifies operational controls such as temperature and flow rate, as well as reactant concentrations of lithium, sodium, calcium, and magnesium. Experts selectively refine input features and output targets based on these insights, ensuring that both machine learning model settings and experimental parameters are meticulously aligned with our objectives to maximize lithium yield and purity.

Building on the "Informed Design and Adjustment" phase, the ML model prediction phase leverages these refined inputs to train machine learning models. These models predict crystallization outcomes, ensuring the configurations within the surrogate space are optimized for maximum lithium yield and purity. Details about the model configurations, training, and operational parameters are provided in section 2.2.4.

In the "Data Acquisition" phase, predictions from the machine learning models guide the generation of a list of potential experiments, as detailed in section 2.2.5. These suggestions, coupled with model predictions, are rigorously reevaluated within the "Data and Results Inspection and Interpretation Modules." Here, statistical analyses help human experts assess the performance of the models and the efficacy of the data acquisition process. During the interpretation phase, the viability of the experiments is evaluated. If the assessments determine that the experiments are feasible the process advances to the experimentation phase. If not, the HITL-AL cycle proceeds with further iterations, adjusting the surrogate space, data acquisition strategies, and model parameters until the experimental plan is finalized.

Once experiments are authorized for "Experimentation," experts review the suggestions, select the most feasible experiments, and conduct them along with reproducibility tests to verify data quality.

This continuous, iterative cycle of review and refinement, depicted in **Figure 1**, enhances the capacity to efficiently produce battery-grade lithium from low-grade brines. Subsequent sections will delve deeper into the experimentation phase and detail the integration of the HITL-AL cycle, illustrating how each step contributes to achieving optimal outcomes.

## 2.1. Experimental Procedures:

The experimental portion of this study consisted of a dual-reactor continuous crystallization setup, inspired by mixed suspension mixed product removal crystallizer (MSMPR) designs used in the pharmaceutical industry [16,35]. The crystallization setup consisted of two, 400 mL reactors set at different temperatures (e.g. 25 °C, and 70 °C respectively; **Figure 2a, b**). Lithium carbonate crystallizes in the higher temperature reactor due to its lower solubility in hot solutions, while crude lithium carbonate solids in the colder reactor slowly dissolve and maintain a steady concentration of lithium carbonate in solution. Liquid was continuously circulated between the two reactors with peristaltic pumps while the flow of solids between reactors was restricted by polyethylene filters slotted into the tubing between the reactors.

At the start of the process, the cold reactor was loaded with the initial solution (e.g., low-grade $Li_2CO_3$ brine). The concentration of various elements in the input solution was monitored before loading the cold reactor. At the end of the crystallization process, the solution was unloaded and filtered from the hot reactor, where the crystallized purified lithium was collected. After unloading the product from the hot reactor, the concentration of different elements in the resulting crystals was measured using Inductively Coupled Plasma Optical Emission Spectroscopy (ICP-OES).

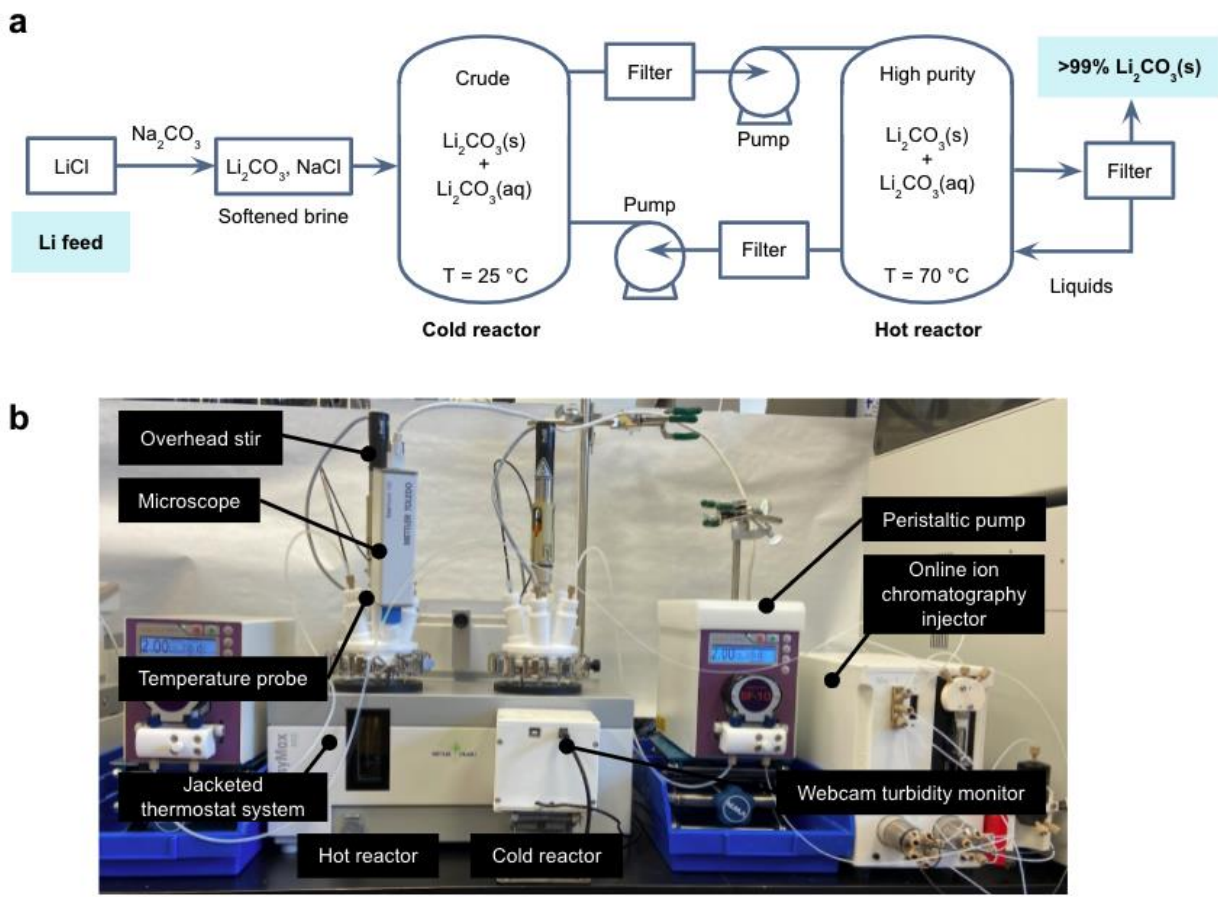

**Figure 2:** (a) Simplified process flow diagram of the continuous hot-cold crystallization setup. (b) Image of benchtop setup used for processing LiCl brines into battery grade $Li_2CO_3$.

The continuous crystallization process was designed to isolate battery-grade $Li_2CO_3$ from a $Li_2CO_3$ impure feedstock; see **Figure 1a** for a schematic overview. The initial crude material for this study was primarily synthesized to simulate various real-world lithium carbonate crudes with typical contaminant levels, including Mg, Ca, K, and Na. To ensure reproducibility, we also used industry-grade feedstock contaminated with a similar set of elements. Control parameters managed within the system included the temperatures of the cold and hot reactors, defining the temperature differential, as well as the pH, stirring rate, slurry concentration, flow rate between reactors, and retention time.

All experiments conducted were systematically recorded in tables and passed to the HITL cycle. Each experiment was assigned a quality score (1, 2, or 3) based on the process expert's observations. A score of 1 indicated no anomalies, such as sedimentation in containers. A score of 2 was given for minor occurrences of such issues, while a score of 3 was assigned to conditions where noticeable anomalies, like sedimentation on container walls, were observed. Additionally, specialists provided comments describing any anomalies noticed during the experimentation.

Throughout various cycles of active learning and experimentation, specialists conducted reproducibility tests. These tests primarily focused on experiments that showed anomalies or unusual results, including a random selection of previous experiments. If contradictory results were observed, the scores could be adjusted. Experiments with significantly divergent outcomes were deemed failed and excluded from further analysis.

## 2.2. Human-in-the-Loop Active Learning Framework:

The HITL-AL framework designed for this study consisted of five iterative steps: Data and Results Inspection, Human Interpretation, Informed Design and Adjustments, Machine Learning Model Prediction, and Data Acquisition. The interactions among these steps were discussed earlier in the opening of Section 2. This section delves deeper, providing more detailed information about each step.

### 2.2.1. Data and Results Inspection:

In the Data and Results Inspection step, experimental (observed) and simulated (ML-predicted surrogate space) data were structured into comprehensive tables, including ICP-OES elemental concentrations results (Ca, K, Li, Mg, Na in ppm) before and after crystallization, alongside operational parameters including reactor temperatures (°C), flow rate (mL/min), and slurry concentration (g/100 mL).

Statistical analyses were systematically performed to explore correlations and identify key parameters influencing process outcomes. Initially, observed experimental data were analyzed to establish baseline relationships. After the first active learning iteration, analyses were extended to incorporate simulated surrogate space conditions and machine learning predictions, thereby elucidating model predictions and highlighting important signals identified by the models.

Pearson correlation analyses quantified linear relationships between input parameters and final impurity levels. SHAP (SHapley Additive exPlanations) values clarified the relative contributions of each feature to model predictions, while sensitivity analyses evaluated the robustness of these predictions by perturbing parameters individually by one standard deviation. The SHAP and sensitivity analyses utilized Random Forest Regressors (RFR) due to their interpretability and straightforward hyperparameter tuning. Hyperparameters—including the number of estimators, maximum depth, minimum samples split, and minimum samples per leaf—were optimized using the Tree-structured Parzen Estimator (TPE) algorithm via Optuna. Collectively, these analyses provided insights into the underlying data relationships and clarified model behaviors.

### 2.2.2. Human Interpretation:

Human experts assessed analytical outcomes guided by structured questions across three key domains: Process Understanding, Impurity Management, and Model Evaluation. For Process Understanding, experts investigated how variations in operational parameters—such as reactor temperatures, flow rates, and slurry concentration—influenced impurity removal. They sought to identify parameter ranges at which these influences became significant. Under Impurity Management, assessments clarified which impurities (e.g., Mg and Ca) were effectively managed under existing experimental conditions and which remained resistant or challenging. In Model Evaluation, experts critically examined whether ML models produced unrealistic or non-physical predictions, such as outputs significantly deviating from observed experimental distributions or predictions lacking expected variability, given the anticipated suboptimal model performance typical of low-data active learning scenarios.

If it was the initial active learning iteration, experts proceeded directly to the informed design and adjustments step. However, in subsequent iterations, experts only moved forward to the informed design and adjustments step when unsatisfactory outcomes or unexpected ML behaviors were identified. These actions included retraining ML models with adjusted hyperparameters, refining surrogate space parameter boundaries, or conducting targeted verification experiments—such as deploying random walkers (described further in section 2.2.3)—to clarify anomalies and potential biases. These iterative refinements, guided by human insights, ensured continuous alignment among model predictions, experimental design, and chemical process understanding.

### 2.2.3. Informed Design and Adjustments:

The surrogate space was constructed using Latin hypercube sampling to uniformly distribute multidimensional combinations of initial elemental concentrations (Ca, K, Li, Mg, Na) and operational parameters (reactor temperatures, flow rates, slurry concentrations). Feature selection and feasible parameter ranges were primarily informed by statistical analyses, including SHAP feature importance scores and sensitivity analysis, alongside operational constraints. Explicit constraints—such as maintaining a minimum temperature differential between reactors ($T_{hot} \geq T_{cold} + 20°C$, later reduced to 2°C, as detailed in the Results section) and normalizing total elemental concentrations to one million ppm—were strictly enforced. When human interpretation indicated insufficient exploration or unexpected predictive anomalies, surrogate space boundaries were adjusted accordingly.

Experts evaluated feasible parameter ranges by identifying biases, model inconsistencies, and persistent trends suggesting areas for further investigation. If surrogate space coverage was deemed inadequate, a random walk algorithm was activated for targeted verification, deploying 100,000 random walkers. Walkers were initialized near flagged regions and took randomized steps within ±25% of identified Pareto frontier boundaries, systematically probing overlooked parameter ranges.

Decisions regarding machine learning model strategies—whether to continue exploratory analyses or transition toward exploitation—were informed by empirical results and experimental observations rather than purely computational predictions. The exploratory analysis phase primarily identified key parameters and their critical ranges necessary for achieving battery-grade outcomes. Once sufficient understanding was reached, the strategy shifted to exploitation, focusing on defining decision boundaries between battery-grade and non-battery-grade conditions, particularly in relation to Mg impurity tolerance. The decision to transition between strategies or initiate targeted verification was systematically informed by human interpretation, ensuring continuous alignment between the experimental workflow and evolving process insights.

### 2.2.4. ML Model Predictions:

Gaussian Process (GP) were selected for their model-agnostic nature, flexibility, and Bayesian properties, providing uncertainty quantification valuable for process optimization in low-data active learning settings [25,36–39]. GP Regressor (GPR) models predicted post-crystallization impurity concentrations, while GP Classifier (GPC) models classified experimental conditions as battery-grade or non-battery-grade. As detailed in Section 2.2.3, input features and prediction targets were selected during the Informed Design and Adjustments step. Thus, GP models were trained using elemental concentrations (Ca, K, Li, Mg, Na) and process control parameters (reactor temperatures, temperature differences, flow rates, slurry concentrations) as input features. GPR models targeted post-refinement impurity concentrations, and GPC models used a binary battery-grade label (True/False) as prediction targets.

Separate GPR models were developed for each target outcome, employing a differentiable Matern kernel, tuned using the Tree-structured Parzen Estimator (TPE) algorithm. Expert evaluations of predictive accuracy informed further hyperparameter refinements and model configurations.

During the initial exploration, GPR models predicted post-crystallization elemental concentrations, aiding experts in identifying key parameters that influence impurity reduction. Battery-grade label predictions using GPC were subsequently incorporated to delineate boundaries between battery-grade and non-battery-grade outcomes, providing insights into how impurity levels impact lithium carbonate purity. Standard scaling was applied to both the training dataset (features only) and surrogate simulation datasets to ensure consistent model performance by maintaining uniform feature magnitudes and preventing numerical inconsistencies in predictions.

### 2.2.5. Data Acquisition:

The role of the data acquisition step is to suggest experiments based on predictions of the ML models. However, it should be noted that these suggestions are not guaranteed to be directly implemented by the experimentalists. In fact, in high dimensional experimental space and low data regimes, sub-optimal predictions are inevitable. Therefore, these suggestions may be adjusted according to the knowledge of experimentation experts.

The data acquisition in this study has two strategies. The first strategy emphasized exploratory analysis in the initial active learning cycles. After gaining some insight an information decision boundary exploitation was followed.

**Exploratory Analysis:** The initial phase involved methodical exploration of the experimental space using GPR predictions of the surrogate space using algorithms like non-dominant sorting of genetic algorithm (NSGA-II), Pareto frontiers were extracted to pinpoint conditions that GPR models believed to minimize key impurities such as calcium and magnesium. These predictions were reviewed by human experts who adjusted experimental parameters to ensure they were practically and theoretically viable.

**Decision Boundary Exploitation**: GPC model was used to classify experimental outcomes as either non-battery-grade or battery-grade. The model calculated the probability of each experimental condition yielding a battery-grade product, with the decision boundary defined at a class probability of 0.5. To interrogate this boundary, a ray-tracing-inspired algorithm was implemented. This algorithm systematically paired each non-battery-grade data point with its nearest battery-grade counterpart to calculate their mathematical midpoint. These midpoints, particularly those located closest to the decision boundary, were prioritized for subsequent experiments to efficiently refine the conditions for producing battery-grade lithium.

## 3. Results and Discussions

Optimization of the continuous lithium crystallization process began with analysis of 16 preliminary experiments conducted before active learning integration. These initial experiments, selected based on expert judgment, consistently produced lithium carbonate ($Li_2CO_3$) with purity exceeding 99%, as shown in **Table S2**.

Post-refinement analyses showed that Na impurities consistently fell below the project's set battery-grade threshold of 500 ppm and for most tests even under the more restrictive limit of 250 ppm, as introduced in **Table S1**, demonstrating this without additional purification treatments. K concentrations, although occasionally exceeding the strict battery-grade threshold of 10 ppm, were deemed manageable due to the high water solubility of potassium salts. Literature indicates typical removal efficiencies of 97%–99% for K through simple multi-stage water washes, allowing reduction from initial concentrations around 500 ppm to within battery-grade specifications (Table S1)[1,2]. Similarly, Ca impurities, despite exceeding the battery-grade threshold of 70 ppm in some preliminary outcomes, were also considered manageable. Literature supports mild supplementary treatments such as dilute acid washes or carbonation–decarbonation recrystallization, which reliably reduce calcium from initial concentrations of a few hundreds to a thousand ppm to below battery-grade limits, typically achieving removal efficiencies between 70%–85% [1,3,4].

Given the satisfactory refinement consistently achieved for Na and Li impurities, further optimization efforts for these elements were deprioritized. Although post-refinement K concentrations occasionally exceeded battery-grade thresholds, optimization for potassium was similarly deprioritized due to the ease and effectiveness of post-process water washing. This decision was corroborated by statistical analyses conducted within the HITL framework. Specifically, SHAP analysis demonstrated that post-refinement K concentrations exhibited lower feature importance than a randomly generated control variable (**Figure S1**). Additionally, Pearson correlation matrices indicated no significant correlations between K concentrations and process control parameters after refinement (**Figure S2**). Although sensitivity analysis suggested a slight influence from flow rate on K reduction, this was statistically insignificant, exhibiting comparable magnitude to random variations (**Figure S1**). Together, these analyses justified assigning K a lower priority for subsequent experimental optimization.

Conversely, Mg impurity removal posed significant challenges. In scenarios where initial Mg concentrations exceeded approximately 80 ppm—the conventional battery-grade threshold—none of the preliminary experiments successfully reduced Mg below this desired limit (**Table S1**). Furthermore, elevated initial Mg concentrations corresponded with increased post-refinement Ca concentrations, occasionally surpassing the battery-grade threshold, though still manageable through previously described post-processing methods. Consequently, informed by initial observations and HITL-driven analyses, subsequent experimental designs specifically prioritized optimizing Mg and Ca removal.

Exploratory analysis began by employing Pareto frontier extraction, aiming to reduce post-refinement Mg and Ca concentrations. Gaussian Process Regression (GPR) predictions were generated across a surrogate experimental space consisting of 10,000 data points, as detailed fully in **Tables S3**. At each exploratory iteration, approximately 30 candidate conditions identified by Pareto frontier analysis were presented to experts, from which a subset was selected for experimental execution. After four active learning iterations, a cumulative total of 36 successful, reproducible experiments—including the initial 16 experiments—were collected for further analysis (**Table S4**). The iterative refinement and optimization achieved through these cycles are illustrated in **Figure 3**.

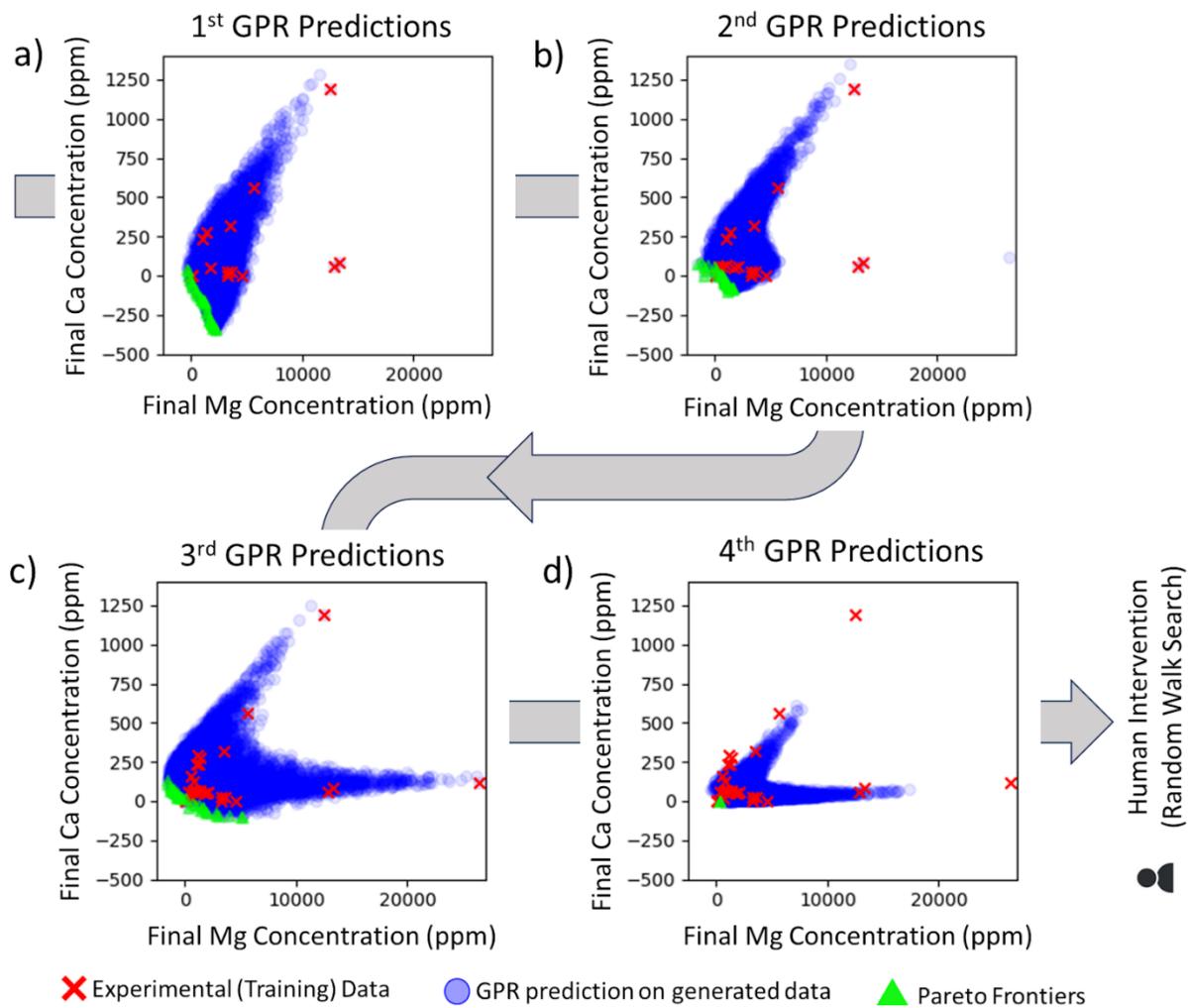

**Figure 3:** GPR predictions and Pareto frontier extraction during the first four cycles of Bayesian active learning interactions on the 10,000 points in the surrogate space.

As depicted in **Figure 3**, despite inherent complexity causing occasional sub-optimal model predictions—including non-physical outcomes such as negative concentrations—GPR predictions remained valuable for identifying meaningful trends and signals.

Throughout the initial exploratory active learning phase, informed by insights from the first 16 expert-designed experiments, it was observed that the continuous crystallization process consistently met battery-grade lithium carbonate specifications for Na (**Table S1** and **S4**). Potassium and calcium impurities were effectively managed with straightforward, cost-efficient secondary treatments, ensuring they did not obstruct overall lithium purification. However, across the first 36 experiments conducted, when initial Mg concentrations exceeded 200 ppm, reducing Mg below the battery-grade threshold of 80 ppm was consistently unattainable (**Table S1** and **S4**). This persistent challenge led to a strategic shift in optimization efforts, concentrating specifically on improving magnesium impurity removal to reliably achieve battery-grade lithium.

To address the persistent challenge of effectively reducing Mg impurities, human experts examined potential biases in the experimental design—specifically concerning the selected parameter ranges. A targeted random walk algorithm was deployed along the boundaries defined by the last identified Pareto frontier, systematically exploring adjacent regions. By allowing deviations up to ±25% from these Pareto frontier boundaries, this approach generated a new surrogate space containing 5,000 additional experimental conditions. GPR models, trained on all experimentally observed data available up to that point (the first 36 data points, **Table S4**), were subsequently employed to predict outcomes across this newly created surrogate space, enabling the detection of overlooked parameter spaces or latent biases.

Analysis of GPR-predicted data from this random walk surrogate space revealed an unexpected inverse correlation between the cold reactor's temperature and predicted post-refinement Mg concentration. This surprising observation, identified through Pearson correlation, SHAP feature importance, and sensitivity analyses performed on the model-predicted dataset (**Figure 4**), starkly contrasted prevailing assumptions derived from prior literature and scientific heuristics. The statistical analyses indicated that higher cold reactor temperatures could significantly enhance Mg impurity removal efficiency—directly challenging established recommendations of maintaining cold reactor temperatures below 60°C with a 20°C differential between reactors. Prompted by these model-derived insights, experimental trials were subsequently proposed and conducted to validate this observation.

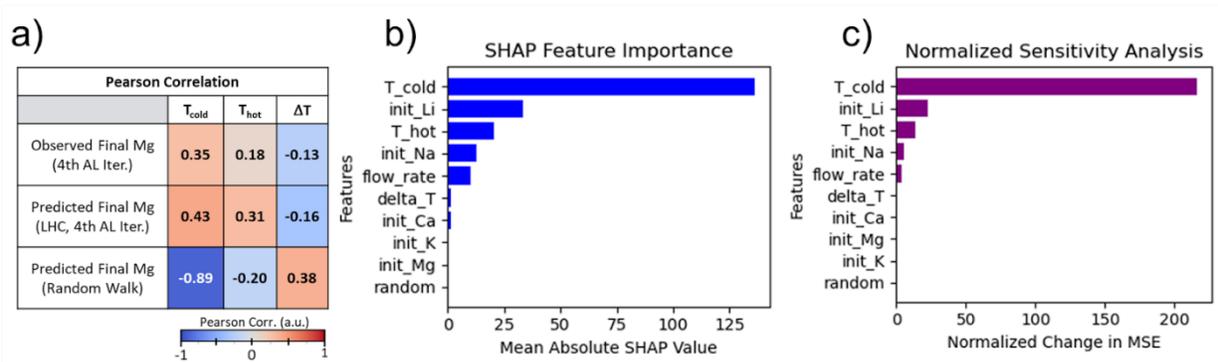

**Figure 4: a)** Pearson correlation analysis illustrating the inverse relationship between cold reactor temperature and magnesium concentration across observed, LHC, and random walk-generated data. **b)** SHAP feature importance analysis revealing the impact of cold reactor temperature on the final Mg concentration. **c)** Normalized sensitivity analyses for the final Mg target from the random walk-generated data, demonstrating the significant influence of cold reactor temperature on reducing final Mg concentrations.

To experimentally validate this correlation, human experts conducted targeted trials, systematically increasing the cold reactor temperature while holding other operational parameters constant. These validation experiments confirmed the initial GPR-derived predictions, substantiating the inverse relationship between cold reactor temperature and post-refinement Mg concentration. **Figure 5** presents one such validation experiment, highlighting the direct impact of cold reactor temperature on Mg impurity reduction (detailed experimental conditions provided in **Table S4** and **S5**).

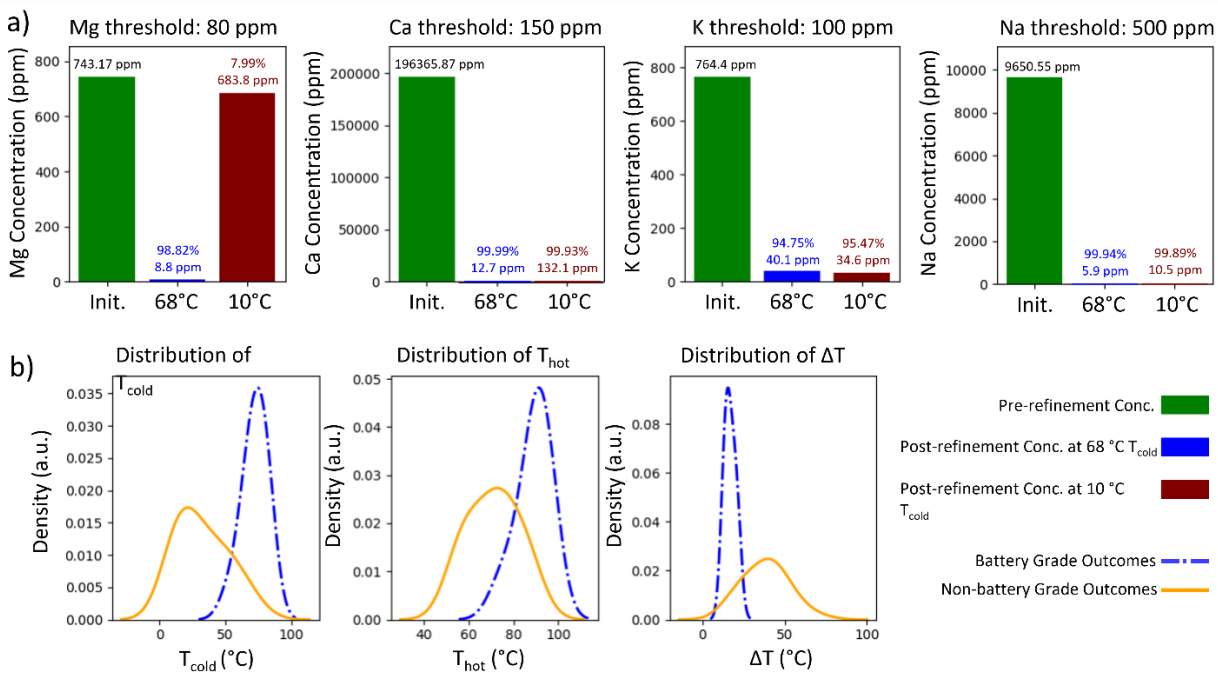

**Figure 5:** An experiment showcasing the impact of reactor temperatures on reducing impurity concentration, especially magnesium concentration (**Table S5**).

Following the experimental validation of the inverse relationship between cold reactor temperature and Mg impurity concentration, the study proceeded to the exploitation phase. The objective in this phase was to clearly delineate the decision boundary between battery-grade and non-battery-grade lithium outcomes, specifically quantifying the maximum permissible initial Mg concentration that could yield battery-grade lithium under optimized temperature conditions. **Figure 6** visualizes this boundary by plotting experimentally observed outcomes—classified as battery-grade or non-battery-grade—against initial Mg concentration and cold reactor temperature. The resulting visualization reveals a distinct and actionable decision boundary.

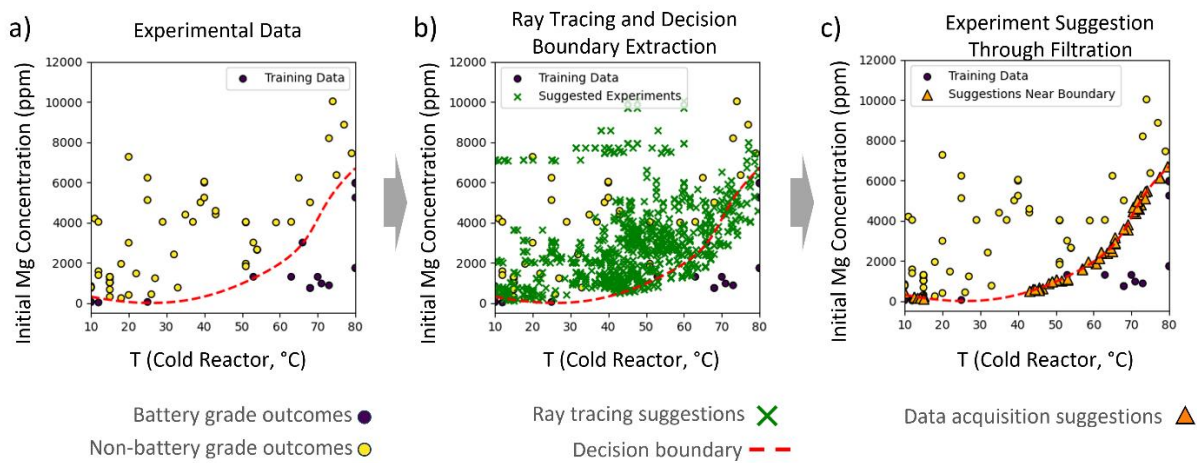

**Figure 6: a)** Experiment refinement outcomes focused on the impact of initial Mg and temperature of cold reactor focused on initial Mg ranges below 12000 ppm. **b)** demonstration of ray tracing and decision boundary extraction process. **c)** Filtered experiments by selecting those closest to the decision boundary.

Subsequently, a GPC model was trained to formalize the experimentally identified decision boundary (**Figure 6**). Utilizing a ray tracing algorithm, experimental candidates with predicted probabilities closest to the critical 0.5 threshold—indicating equal likelihood of achieving battery-grade or non-battery-grade lithium—were selected for further verification experiments.

**Figure 7** presents these GPR-derived predictions, distinguishing battery-grade from non-battery-grade outcomes, overlaid with all observed experimental data (totaling 80 experiments). This visualization highlights a substantial improvement in permissible initial Mg contamination levels. Historically, industry standards limited acceptable Mg contamination to approximately 80 ppm. However, our HITL-driven optimization demonstrates that increasing the cold reactor temperature beyond the initially recommended limit of 60°C can significantly elevate the tolerance of initial Mg contamination to several thousand ppm. This finding implies a potential reduction in the extent of pre-refinement processes and lessens reliance on higher-purity lithium sources to achieve battery-grade products.

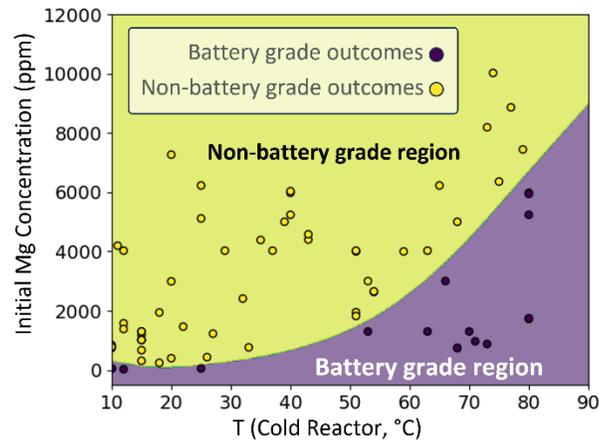

**Figure 7:** Projection of GPC model predictions on the initial Mg concentration and cold reactor temperature plane, illustrating the decision boundary between battery-grade and non-battery-grade outcomes. This visualization emphasizes improved Mg contamination tolerance, demonstrating that higher initial impurity levels can still yield battery-grade lithium by increasing the temperature of the cold reactor.

**Figure 8** provides additional analytical evidence supporting the significant influence of cold reactor temperature and initial Mg concentration on process outcomes. Specifically, **Figure 8.a** shows the distribution of cold reactor temperatures for experiments with initial Mg concentrations above 200 ppm, categorized by battery-grade or non-battery-grade outcomes. This distribution clearly demonstrates that higher reactor temperatures strongly correlate with successful Mg impurity reduction. Further supporting these observations, **Figure 8.b** presents SHAP analysis results, highlighting the prominent impact of cold reactor temperature on achieving battery-grade lithium. Lastly, **Figure 8.c** illustrates sensitivity analysis outcomes, reinforcing that both initial Mg concentration and cold reactor temperature are key parameters determining the purity of the final lithium carbonate product.

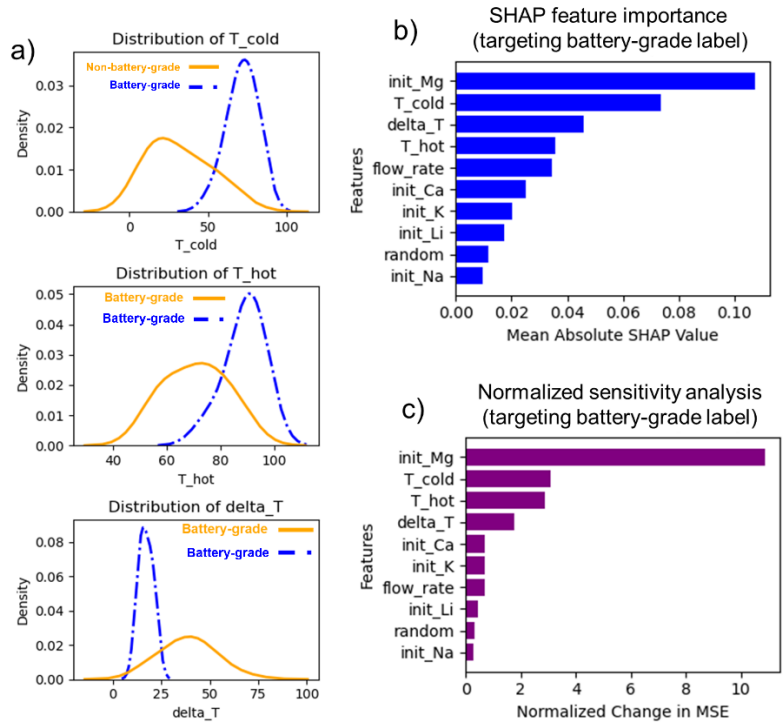

**Figure 8**: **a)** shows the kernel density estimate distribution of reactor temperatures against initial magnesium concentrations. **b)** presents SHAP analysis indicating the influence of temperature on achieving battery-grade outcomes. **c)** illustrates sensitivity analysis results, highlighting critical temperature-related parameters.

Including the initial 16 historic experiments, a total of 38 experiments (22 additional experiments) were required to obtain clear experimental evidence of the influence of cold reactor temperature on achieving battery-grade lithium carbonate. In total, 80 experiments were conducted throughout the study to validate these insights and precisely identify the decision boundary between battery-grade and non-battery-grade outcomes (as represented in **Figures 6-8**). Given the complexity of the crystallization process and the extensive dimensionality of the parameter space, identifying critical parameters within just 38 experiments demonstrates the effectiveness of integrating human expertise, statistical analyses, and machine learning-driven active learning methods. Conducted at an accelerated pace of approximately one experiment per day, this human-in-the-loop approach substantially reduced the timeline traditionally required for such process optimization, facilitating rapid decision-making and iterative refinement despite the challenges inherent in low-data machine learning scenarios.

To further illustrate how the HITL framework accelerated the identification of experimental conditions conducive to producing battery-grade lithium carbonate with initial Mg concentrations exceeding 200 ppm, we compared its efficacy against two alternative active learning frameworks without human intervention. Specifically, we compared the performance of our HITL-assisted active learning (HITL-AL) approach, which required 38 experiments to identify key conditions leading to battery-grade outcomes, against the performance of purely computational active learning methods without expert guidance. These methods included a random-sampling approach and a simplified Bayesian optimization approach, both conducting one experiment per active learning cycle.

For this comparative evaluation, a surrogate GPC model—trained on data and insights obtained from the HITL approach—was used to predict battery-grade outcomes. Two distinct experimental datasets, each comprising 10,000,000 simulated experimental scenarios, were generated for this comparison. The first dataset, termed "uninformed," strictly adhered to initial parameter ranges without incorporating insights from the HITL process. The second dataset, termed "informed," was explicitly constrained by parameter ranges informed by the HITL-identified optimal temperature settings and impurity conditions.

Each active learning method underwent 100 computational simulations (instances) employing different random seeds to statistically evaluate the frequency at which battery-grade conditions were identified. The first active learning approach applied a random sampling data acquisition policy. In contrast, the second method used a simplified Bayesian approach explicitly adapted for classification tasks, predicting experimental outcomes as either battery-grade or non-battery-grade. The classification approach was adopted due to challenges in reliably simulating detailed regression predictions of impurity concentrations with the surrogate GPC model. For condition selection, the Bayesian method relied on the upper confidence bound (UCB) metric to target experimental conditions with high uncertainty.

**Figure 9** summarizes the comparative results of the simplified Bayesian and random active learning methods using both informed and uninformed datasets. The simplified Bayesian approach leveraging the informed dataset (integrating optimized parameter ranges derived from prior HITL insights) identified battery-grade conditions within 40 experiments at a success rate of approximately 67%, significantly outperforming the uninformed dataset scenario, which achieved only 14% success. Similarly, the random sampling approach achieved success rates of 14% with the informed dataset and only 1% with the uninformed dataset. These results clearly illustrate the substantial advantage provided by incorporating human-derived knowledge and insights into active learning optimization strategies.

Collectively, these findings underscore the practical effectiveness and importance of integrating human expertise with machine learning analyses. Targeted human interventions, guided by nuanced insights from ML-driven data exploration, notably accelerated the discovery of optimal process parameters, demonstrating a highly effective synergy for refining complex chemical systems.

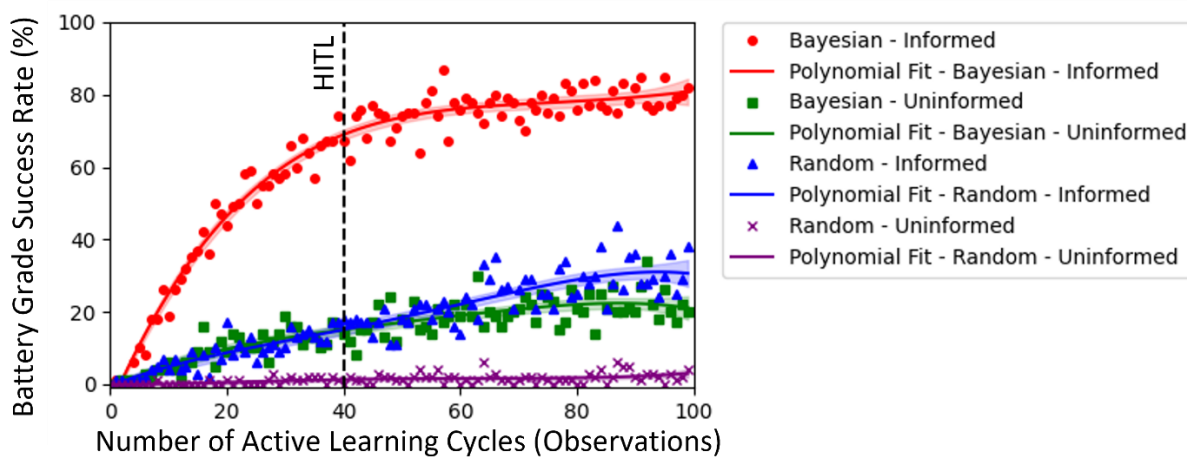

**Figure 9:** Comparison of success rates in identifying battery-grade conditions using HITL active learning, human-independent Bayesian active learning, and random sampling methods, with and without prior information.

## Conclusions

In this study, we presented a Human-in-the-Loop (HITL)-assisted active learning framework tailored specifically for the optimization of continuous lithium carbonate crystallization processes from low-grade brine sources, such as those found in the Smackover Formation. By seamlessly integrating human domain expertise with artificial intelligence-driven methods, we effectively addressed critical challenges associated with complex impurity management, especially regarding magnesium, which has historically constrained battery-grade lithium production.

Through iterative experimentation and informed interpretation of machine learning model predictions, our approach notably accelerated the identification of optimal process conditions. A critical insight emerged from this collaborative optimization: contrary to traditional assumptions recommending lower cold reactor temperatures (below 60°C with a 20°C temperature differential), our framework identified and experimentally validated that elevated cold reactor temperatures significantly enhanced magnesium impurity removal efficiency. This finding permitted an unprecedented increase in magnesium impurity tolerance—from conventional limits of approximately 80 ppm up to several thousand ppm—markedly reducing the need for intensive pre-refinement stages.

Overall, this research demonstrates the pivotal role human expertise can play in refining AI-driven optimization processes, particularly in high-dimensional chemical systems constrained by limited experimental throughput. By effectively balancing human intuition, domain knowledge, and machine learning analysis, our approach not only expedited critical discoveries but also facilitated agile responses to emergent insights, thereby optimizing experimental efficiency and accuracy. Consequently, the methodological and practical advancements presented herein represent significant progress toward the sustainable, economically viable extraction of lithium carbonate from complex, impurity-rich brines. This advancement contributes directly to the broader objective of responsibly harnessing North America's abundant but challenging lithium resources, supporting the growth and sustainability of lithium-based technologies vital for the global energy transition.

## Data availability

The experimental data supporting this article are detailed in Table S4 of the supplementary material, which also includes additional information on experimental procedures and complementary analyses. All experimental data, AI-generated experiment suggestions, executable Python scripts, and resources necessary for reproducing the paper's figures and analyses are hosted on our dedicated GitHub repository. For full access to these materials and more details, please visit our GitHub Repository at https://github.com/shmouses/HITL_Adaptive_Active_Learning_Lithium .

## Author Information

Corresponding Author:

Jason Hattrick-Simpers – Canmet MATERIALS, Natural Resources Canada, 183 Longwood Rd S, Hamilton, ON, Canada; orcid.org/0000-0003-2937-3188

Email: jason.hattrick-simpers@nrcan-rncan.gc.ca

Authors:

S. Shayan Mousavi Masouleh – Canmet MATERIALS, Natural Resources Canada, 183 Longwood Rd S, Hamilton, ON, Canada; Department of Materials Science and Engineering, McMaster University, 1280 Main St W, Hamilton, ON, Canada; Clean Energy Innovation, National Research Council of Canada, 2620 Speakman Dr, Mississauga, ON, Canada; orcid.org/0000-0003-0313-7590

Corey A. Sanz – Telescope Innovations, 301-2386 E Mall, Vancouver, BC, Canada; orcid.org/0000-0002-3836-0744

Ryan P. Jansonius – Telescope Innovations, 301-2386 E Mall, Vancouver, BC, Canada; orcid.org/0000-0002-4014-2068

Cara Cronin – Telescope Innovations, 301-2386 E Mall, Vancouver, BC, Canada;

Jason E. Hein – Telescope Innovations, 301-2386 E Mall, Vancouver, BC, Canada; orcid.org/0000-0002-4345-3005

**Notes**

The authors declare that there are no competing financial interests or conflicts of interest.

**Acknowledgment**

The authors gratefully acknowledge funding from the Critical Minerals Research, Development, and Demonstration (CMRDD) Program administered by Natural Resources Canada. We also extend our gratitude to Standard Lithium for their financial support and valuable discussions that have guided the project's design. Additionally, Telescope Innovations thanks the Mining Innovation Commercialization Accelerator (MICA) for financial support related to this project. We sincerely appreciate the reviewers for their constructive feedback, which has contributed to the refinement of the manuscript.

# Human-AI Synergy in Adaptive Active Learning for Continuous Lithium Carbonate Crystallization Optimization

Shayan Mousavi Masouleh [1,2] Corey A. Sanz [3] Ryan P. Jansonius [3] Cara Cronin [3] Jason E. Hein [3] Jason Hattrick-Simpers [1] *

[1] Canmet MATERIALS Natural Resources Canada 183 Longwood Rd S Hamilton ON Canada

[2] Clean Energy Innovation National Research Council of Canada 2620 Speakman Dr Mississauga ON Canada

[3] Telescope Innovations 301-2386 E Mall Vancouver BC Canada

**Corresponding Author**

* jason.hattrick-simpers@nrcan-rncan.gc.ca

## Supplementary Information:

**Table S1:** Comparative specifications for battery-grade lithium carbonate. In the absence of a universal ISO standard the target parameters for this study are compared against established industry and national standards (1–3).

| Parameter | This Study | China YS/T 582-2006 | China YS/T 582-2013 | Arcadium Lithium (Livent-2022) |
|---|---|---|---|---|
| Grade | Battery Grade | Battery Grade | Battery Grade | Micronized Battery Grade |
| $Li_2CO_3$ (%) | ≥ 99.5 | ≥ 99.5 | ≥ 99.5 | ≥ 99.5 |
| Na (ppm) | ≤ 500 | ≤ 300 | ≤ 250 | ≤ 500 |
| Mg (ppm) | ≤ 80 | ≤ 100 | ≤ 80 | N/A |
| Ca (ppm) | ≤ 150 | ≤ 70 | ≤ 50 | ≤ 400 |
| K (ppm) | ≤ 100 | ≤ 20 | ≤ 10 | N/A |
| Si (ppm) | ≤ 30 | N/A | ≤ 30 | N/A |
| Primary Measurement Method(s) | ICP-OES | FAAS, Titration, Gravimetry | ICP-OES, Titration, Ion Chromatography | ICP-OES, Titration |

**Table S2:** Initial dataset comprising 16 continuous lithium recrystallization experiments (expert-designed experiments). Green cells indicate conditions within battery-grade specifications. Yellow cells denote conditions slightly exceeding battery-grade thresholds but not critically problematic. Red cells identify conditions significantly beyond battery-grade limits. Orange cells highlight initial brine concentrations considered as highly impure.

| Exp. # | T cold (°C) | T hot (°C) | T difference (°C) | flow rate (mL/min) | slurry concentration (g total solid/100 mL) | init_Ca (ppm) | init_K (ppm) | init_Li (ppm) | init_Mg (ppm) | init_Na (ppm) | initi_Li$_2$CO$_3$ purity (%) | fini_Ca (ppm) | fini_K (ppm) | fini_Li (ppm) | fini_Mg (ppm) | fini_Na (ppm) | fini_Li$_2$CO$_3$ purity (%) |
|---|---|---|---|---|---|---|---|---|---|---|---|---|---|---|---|---|---|
| 1 | 25 | 70 | 45 | 2.00 | 2.00 | 0.00 | 67.60 | 172237.60 | 51.50 | 109.00 | 99.87% | 0.00 | 0.00 | 190934.90 | 71.60 | 41.10 | 99.93% |
| 2 | 25 | 70 | 45 | 1.40 | 2.00 | 47253.80 | 2756.70 | 135952.20 | 5113.40 | 1034.00 | 70.76% | 231.80 | 19.60 | 187466.80 | 992.50 | 59.90 | 98.99% |
| 3 | 10 | 80 | 70 | 5.25 | 23.00 | 0.00 | 67.60 | 172237.60 | 51.50 | 109.00 | 99.87% | 0.00 | 0.00 | 192859.80 | 62.10 | 29.90 | 99.85% |
| 4 | 15 | 65 | 50 | 7.20 | 4.03 | 133045.20 | 2819.00 | 110257.10 | 14152.20 | 8741.60 | 40.98% | 321.60 | 0.00 | 190541.00 | 3463.00 | 19.40 | 97.40% |
| 5 | 10 | 74 | 64 | 1.00 | 6.00 | 0.00 | 67.60 | 172237.60 | 51.50 | 109.00 | 99.87% | 5.00 | 15.90 | 182949.60 | 6.90 | 3.00 | 99.96% |
| 6 | 10 | 79 | 69 | 2.90 | 4.88 | 133045.20 | 2819.00 | 110257.10 | 14152.20 | 8741.60 | 40.98% | 563.00 | 0.00 | 184390.70 | 5589.70 | 13.50 | 95.92% |
| 7 | 10 | 50 | 40 | 4.00 | 12.54 | 0.00 | 67.60 | 172237.60 | 51.50 | 109.00 | 99.87% | 5.00 | 15.10 | 185909.60 | 1.00 | 2.00 | 99.91% |
| 8 | 10 | 80 | 70 | 1.20 | 2.98 | 133045.20 | 2819.00 | 110257.10 | 14152.20 | 8741.60 | 40.98% | 1191.10 | 95.60 | 175798.80 | 12472.40 | 14.60 | 91.34% |
| 9 | 40 | 60 | 20 | 2.00 | 2.47 | 133045.20 | 2819.00 | 110257.10 | 14152.20 | 8741.60 | 40.98% | 0.00 | 85.50 | 182718.40 | 3236.10 | 9.00 | 97.36% |
| 10 | 35 | 58 | 23 | 4.00 | 2.01 | 57351.90 | 2409.50 | 157345.80 | 4387.20 | 2642.20 | 70.20% | 0.00 | 79.30 | 183864.10 | 4528.10 | 9.50 | 96.72% |
| 11 | 26 | 74 | 48 | 0.50 | 9.99 | 90357.40 | 2703.10 | 107587.90 | 49857.80 | 8578.40 | 41.52% | 81.80 | 632.10 | 169606.00 | 13326.70 | 416.30 | 90.50% |
| 12 | 30 | 85 | 55 | 6.00 | 12.01 | 198054.90 | 7813.00 | 79531.90 | 26040.20 | 5626.80 | 25.08% | 22.30 | 94.60 | 186056.10 | 3220.20 | 12.20 | 97.66% |
| 13 | 20 | 60 | 40 | 4.50 | 8.05 | 25383.10 | 2109.10 | 172910.40 | 7265.20 | 3645.30 | 81.82% | 26.50 | 92.70 | 185493.60 | 3639.90 | 9.90 | 97.85% |
| 14 | 32 | 76 | 44 | 1.50 | 7.43 | 77578.60 | 3988.30 | 96916.80 | 350.20 | 95148.80 | 35.37% | 276.80 | 187.90 | 183972.80 | 1421.80 | 363.60 | 98.67% |
| 15 | 25 | 80 | 55 | 3.00 | 4.64 | 9820.60 | 2069.80 | 99184.70 | 88738.00 | 4776.30 | 48.48% | 62.20 | 77.10 | 180032.60 | 12911.20 | 30.70 | 90.87% |
| 16 | 11 | 52 | 41 | 3.60 | 7.67 | 52767.80 | 38502.50 | 130256.00 | 4194.10 | 31253.60 | 50.69% | 52.40 | 77.00 | 187044.70 | 1723.70 | 39.00 | 98.95% |

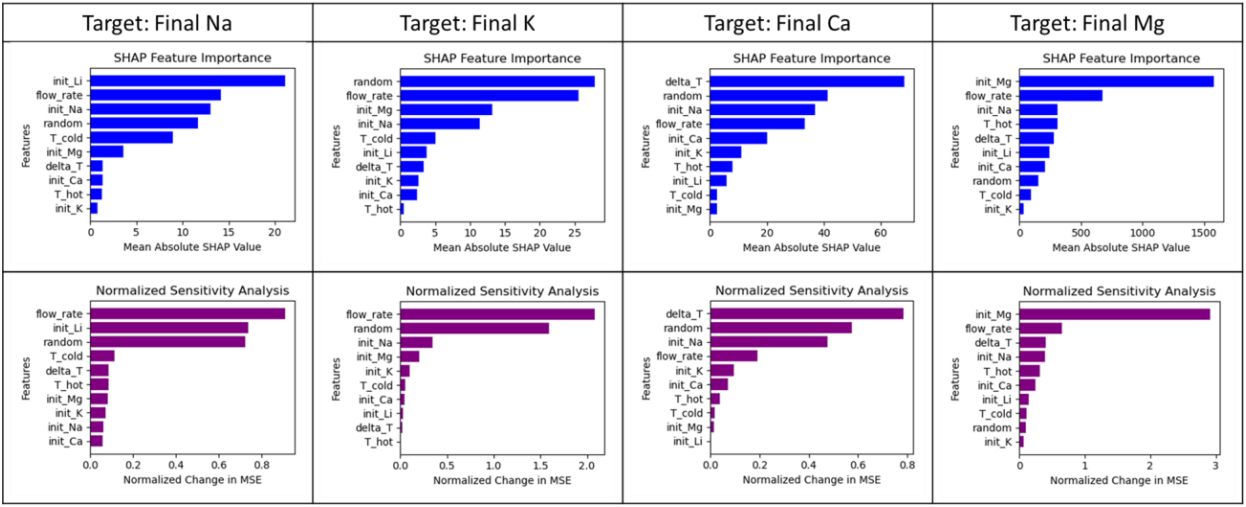

**Figure S1:** SHAP feature importance and normalized sensitivity analyses performed on the initial set of 16 continuous lithium recrystallization experiments (expert-designed experiments).

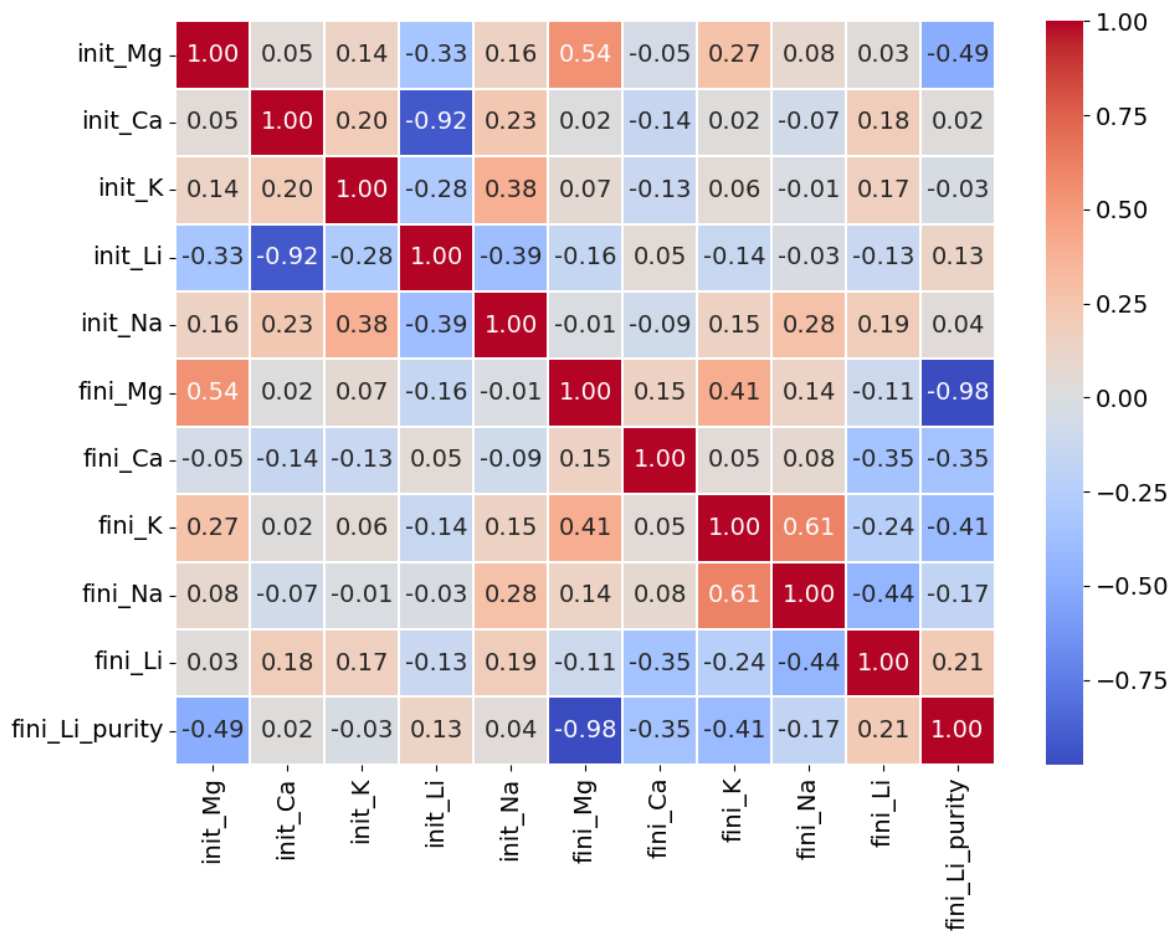

**Figure S2:** Pearson correlation matrix for iteration 0 analyzing relationships among input parameters initial impurity concentrations and post-refinement impurity concentrations from the initial set of 16 expert-designed experiments.

**Table S3:** List of parameter ranges for all the major surrogate data spaces simulated in this study using LHC sampling. Surrogate space A with 10000 was the majorly used space in exploration phase.

| Label | # Points | T_cold Min (°C) | T_cold Max (°C) | T_hot Min (°C) | T_hot Max (°C) | flow_rate Min (mL/min) | flow_rate Max (mL/min) | init_Ca Min (ppm) | init_Ca Max (ppm) | init_K Min (ppm) | init_K Max (ppm) | init_Li Min (ppm) | init_Li Max (ppm) | init_Mg Min (ppm) | init_Mg Max (ppm) | init_Na Min (ppm) | init_Na Max (ppm) | slurry_concentration Min (g total solid/100 mL) | slurry_concentration Max (g total solid/100 mL) |
|---|---|---|---|---|---|---|---|---|---|---|---|---|---|---|---|---|---|---|---|
| A | 10000 | 10 | 60 | 40 | 80 | 0.5 | 6.0 | 20000 | 200000 | 500 | 2000 | 20000 | 170000 | 100 | 1000 | 500 | 15000 | 1.5 | 10 |
| B | 50000 | 40 | 90 | 42 | 95 | 0.5 | 6.0 | 5000 | 260000 | 500 | 10000 | 20000 | 170000 | 100 | 50000 | 500 | 25000 | 1.5 | 10 |
| C | 100000 | 10 | 90 | 15 | 95 | 0.5 | 6.0 | 5000 | 200000 | 500 | 10000 | 20000 | 170000 | 100 | 50000 | 500 | 15000 | 1.5 | 10 |
| D | 100000 | 40 | 90 | 45 | 95 | 2.0 | 6.0 | 2500 | 300000 | 250 | 15000 | 20000 | 185000 | 50 | 75000 | 250 | 22500 | 1.5 | 10 |
| E | 150000 | 30 | 85 | 35 | 90 | 1.0 | 7.0 | 10000 | 200000 | 500 | 10000 | 20000 | 170000 | 50 | 12000 | 500 | 15000 | 1.5 | 10 |
| F | 400000 | 60 | 90 | 65 | 95 | 2.0 | 5.0 | 30000 | 200000 | 7000 | 850000 | 20000 | 150000 | 1000 | 5000 | 22000 | 88000 | 1.5 | 10 |

**Table S4:** Comprehensive overview of experimental data. Experiments labeled Exp.# 1-16 were expert-designed (also shown in **Table S2**), while the 4th HITL-AL cycle concludes at Exp.# 36. The remaining data were acquired during the random walk process and decision boundary exploitation between battery-grade and non-battery-grade outcomes.

| Exp. # | T cold (°C) | T hot (°C) | ΔT (°C) | Flow rate (mL/min) | slurry concentration (g total solid per 100 mL) | Initial Ca (ppm) | Initial K (ppm) | Initial Li (ppm) | Initial Mg (ppm) | Initial Na (ppm) | Initial Li2CO3 purity (%) | Final Ca (ppm) | Final K (ppm) | Final Li (ppm) | Final Mg (ppm) | Final Na (ppm) | Final Li2CO3 purity (%) |
|---|---|---|---|---|---|---|---|---|---|---|---|---|---|---|---|---|---|
| 1 | 25 | 70 | 45 | 2.0 | 2.00 | 0.00 | 67.60 | 172237.60 | 51.50 | 109.00 | 99.87% | 0.00 | 0.00 | 190934.90 | 71.60 | 41.10 | 99.93% |
| 2 | 25 | 70 | 45 | 1.0 | 2.00 | 47253.80 | 2756.70 | 135952.20 | 5113.40 | 1034.00 | 70.76% | 231.80 | 19.60 | 187466.80 | 992.50 | 59.90 | 98.99% |
| 3 | 10 | 80 | 70 | 5.0 | 23.00 | 0.00 | 67.60 | 172237.60 | 51.50 | 109.00 | 99.87% | 0.00 | 0.00 | 192859.80 | 62.10 | 29.90 | 99.85% |
| 4 | 15 | 65 | 50 | 7.0 | 4.03 | 133045.20 | 2819.00 | 110257.10 | 14152.20 | 8741.60 | 40.98% | 321.6 | 0.00 | 190541.00 | 3463.00 | 19.40 | 97.40% |
| 5 | 10 | 74 | 64 | 1.0 | 6.00 | 0.00 | 67.60 | 172237.60 | 51.50 | 109.00 | 99.87% | 5.00 | 15.90 | 182949.60 | 6.90 | 3.00 | 99.96% |
| 6 | 10 | 79 | 69 | 3.0 | 4.88 | 133045.20 | 2819.00 | 110257.10 | 14152.20 | 8741.60 | 40.98% | 563 | 0 | 184390.70 | 5589.70 | 13.50 | 95.92% |
| 7 | 10 | 50 | 40 | 4.0 | 12.54 | 0.00 | 67.60 | 172237.60 | 51.50 | 109.00 | 99.87% | 5.00 | 15.10 | 185909.60 | 1.00 | 2.00 | 99.91% |
| 8 | 10 | 80 | 70 | 1.0 | 2.98 | 133045.20 | 2819.00 | 110257.10 | 14152.20 | 8741.60 | 40.98% | 1191.10 | 95.60 | 175798.80 | 12472.40 | 14.60 | 91.34% |
| 9 | 40 | 60 | 20 | 2.0 | 2.47 | 133045.20 | 2819.00 | 110257.10 | 14152.20 | 8741.60 | 40.98% | 0.00 | 85.50 | 182718.40 | 3236.10 | 9.00 | 97.36% |
| 10 | 35 | 58 | 23 | 4.0 | 2.01 | 57351.90 | 2409.50 | 157345.80 | 4387.20 | 2642.20 | 70.20% | 0.00 | 79.30 | 183864.10 | 4528.10 | 9.50 | 96.72% |
| 11 | 26 | 74 | 48 | 1.0 | 9.99 | 90357.40 | 2703.10 | 107587.90 | 49857.80 | 8578.40 | 41.52% | 81.80 | 632.10 | 169606.00 | 13326.70 | 416.30 | 90.50% |
| 12 | 30 | 85 | 55 | 6.0 | 12.01 | 198054.90 | 7813.00 | 79531.90 | 26040.20 | 5626.80 | 25.08% | 22.30 | 94.60 | 186056.10 | 3220.20 | 12.20 | 97.66% |
| 13 | 20 | 60.0 | 40 | 5.0 | 8.05 | 25383.10 | 2109.10 | 172910.40 | 7265.20 | 3645.30 | 81.82% | 26.50 | 92.70 | 185493.60 | 3639.90 | 9.90 | 97.85% |
| 14 | 32 | 76.0 | 44 | 2.0 | 7.43 | 77578.60 | 3988.30 | 96916.80 | 350.20 | 95148.80 | 35.37% | 276.80 | 187.90 | 183972.80 | 1421.80 | 363.60 | 98.67% |
| 15 | 25 | 80.0 | 55 | 3.0 | 4.64 | 9820.60 | 2069.80 | 99184.70 | 88738.00 | 4776.30 | 48.48% | 62.20 | 77.10 | 180032.60 | 12911.20 | 30.70 | 90.87% |
| 16 | 11 | 52.0 | 41 | 4.0 | 7.67 | 52767.80 | 38502.50 | 130256.00 | 4194.10 | 31253.60 | 50.69% | 52.40 | 77.00 | 187044.70 | 1723.70 | 39.00 | 98.95% |
| 17 | 33 | 81 | 48 | 5.0 | 14.59 | 73704.50 | 509.60 | 153654.00 | 762.80 | 6234.20 | 65.42% | 63.20 | 74.90 | 183377.30 | 1072.60 | 27.80 | 99.03% |
| 18 | 18 | 54 | 36 | 3.0 | 5.25 | 59899.00 | 2077.40 | 155575.80 | 1941.40 | 52.10 | 70.86% | 59.30 | 52.00 | 185460.80 | 2063.60 | 9.10 | 98.59% |
| 19 | 43 | 70 | 27 | 1.0 | 11.83 | 34719.10 | 1232.90 | 162623.20 | 8837.80 | 72.50 | 78.38% | 115.70 | 143.50 | 176638.50 | 26478.50 | 61.20 | 81.43% |
| 20 | 20 | 75 | 55 | 5.0 | 4.20 | 10641.20 | 124.20 | 183652.30 | 397.20 | 1975.20 | 92.78% | 59.40 | 95.20 | 188364.40 | 287.40 | 47.10 | 99.64% |
| 21 | 27 | 66 | 39 | 3.0 | 6.07 | 31813.10 | 109.70 | 169015.50 | 1227.00 | 3639.60 | 80.84% | 56.60 | 122.10 | 187259.40 | 688.00 | 50.60 | 99.44% |
| 22 | 22 | 77 | 55 | 2.0 | 3.05 | 48229.30 | 60.80 | 159169.30 | 1464.20 | 4920.50 | 72.21% | 129.00 | 36.60 | 178459.80 | 784.30 | 0.00 | 99.23% |
| 23 | 15 | 57 | 42 | 3.0 | 2.65 | 38775.10 | 59.90 | 167015.70 | 1179.30 | 3468.10 | 76.79% | 62.70 | 28.70 | 181291.30 | 827.60 | 0.00 | 99.43% |
| 24 | 32 | 80 | 48 | 2.0 | 1.53 | 25457.90 | 54.50 | 173159.50 | 2410.70 | 709.50 | 83.70% | 46.70 | 36.90 | 179925.90 | 1959.70 | 0.00 | 98.33% |
| 25 | 12 | 61 | 49 | 4.0 | 1.99 | 16021.70 | 74.90 | 175489.90 | 1574.20 | 6102.30 | 86.45% | 236.30 | 43.20 | 185154.30 | 1208.70 | 0.00 | 98.86% |

| | | | | | | | | | | | | | | | | |
|---|---|---|---|---|---|---|---|---|---|---|---|---|---|---|---|---|
| 26 | 26 | 71 | 45 | 3.0 | 13.11 | 33816.80 | 31.50 | 167726.50 | 433.30 | 3072.20 | 81.78% | 16.30 | 0.00 | 173970.30 | 625.70 | 60.20 | 99.48% |
| 27 | 15 | 77 | 62 | 3.0 | 10.63 | 35170.20 | 350.20 | 165793.60 | 310.50 | 6056.80 | 79.83% | 158.20 | 0.00 | 175709.10 | 484.80 | 0.00 | 99.52% |
| 28 | 12 | 61 | 49 | 4.0 | 2.00 | 15082.00 | 9.30 | 166358.00 | 1383.80 | 1015.80 | 89.09% | 296.90 | 0.00 | 179444.80 | 1188.60 | 0.00 | 99.07% |
| 29 | 24 | 66 | 42 | 1.0 | 8.20 | 43599.80 | 135.80 | 159930.10 | 221.90 | 2644.60 | 77.43% | 69.10 | 0.00 | 173457.80 | 1340.20 | 0.00 | 98.83% |
| 30 | 10 | 74 | 64 | 5.0 | 9.93 | 40561.10 | 0.00 | 157745.10 | 844.50 | 2167.00 | 71.72% | 39.00 | 55.00 | 180544.40 | 542.00 | 8.00 | 99.64% |
| 32 | 10 | 74 | 64 | 1.0 | 5.99 | 53.77 | 233.02 | 197915.05 | 0.00 | 44.81 | 99.83% | 5.90 | 59.90 | 179539.00 | 3.90 | 5.90 | 99.95% |
| 33 | 10 | 50 | 40 | 4.0 | 12.55 | 85.61 | 256.84 | 188284.91 | 32.10 | 32.10 | 99.78% | 16.70 | 50.10 | 183303.60 | 19.80 | 3.10 | 99.94% |
| 34 | 15 | 57 | 42 | 3.0 | 2.64 | 23723.43 | 201.50 | 178440.07 | 668.12 | 1781.64 | 85.05% | 52.00 | 44.70 | 185408.10 | 1510.80 | 4.20 | 99.12% |
| 35 | 15 | 57 | 42 | 3.0 | 2.65 | 34279.75 | 170.86 | 168718.32 | 1014.51 | 1356.24 | 77.23% | 35.80 | 48.10 | 179626.70 | 1497.90 | 5.10 | 99.12% |
| 36 | 15 | 57 | 42 | 3.0 | 2.67 | 34279.75 | 170.86 | 168718.32 | 1014.51 | 1356.24 | 77.23% | 19.90 | 42.90 | 177721.90 | 1434.30 | 2.10 | 99.16% |
| 37 | 15 | 57.0 | 42 | 3.0 | 2.66 | 34279.75 | 170.86 | 168718.32 | 1014.51 | 1356.24 | 77.23% | 25.10 | 40.10 | 179616.10 | 1552.70 | 4.00 | 99.10% |
| 38 | 68 | 89 | 21 | 5.0 | 6.08 | 196365.87 | 764.40 | 84126.47 | 743.17 | 9650.55 | 28.84% | 12.70 | 40.10 | 185052.10 | 8.80 | 5.90 | 99.94% |
| 39 | 10 | 83 | 73 | 5.0 | 6.12 | 196365.87 | 764.40 | 84126.47 | 743.17 | 9650.55 | 28.84% | 132.10 | 34.60 | 186958.60 | 683.80 | 10.50 | 99.52% |
| 40 | 68 | 89 | 21 | 5.0 | 6.09 | 196365.87 | 764.40 | 84126.47 | 743.17 | 9650.55 | 28.84% | 0.00 | 39.00 | 182100.80 | 3.20 | 4.20 | 99.90% |
| 41 | 73 | 86 | 13 | 6.0 | 6.38 | 238619.40 | 293.10 | 77592.00 | 877.5 | 4713.80 | 24.09% | 35.2 | 32.3 | 188603.50 | 8.80 | 5.90 | 99.91% |
| 42 | 10 | 60 | 50 | 1.0 | 5.76 | 57449.50 | 2581.00 | 146488.20 | 778.60 | 5143.80 | 68.95% | 22.70 | 36.50 | 179122.30 | 671.60 | 2.00 | 99.46% |
| 43 | 71 | 90 | 19 | 5.0 | 7.10 | 238163.80 | 1172.30 | 67736.50 | 967.50 | 7863.70 | 21.44% | 5.30 | 25.40 | 184055.10 | 54.10 | 0.00 | 99.90% |
| 44 | 15 | 41 | 26 | 2.0 | 2.81 | 125946.60 | 1257.20 | 117827.50 | 730.10 | 9209.70 | 46.21% | 182.60 | 23.10 | 185158.20 | 189.90 | 8.40 | 99.75% |
| 45 | 66 | 87 | 21 | 5.0 | 3.31 | 10000.00 | 4000.00 | 174796.00 | 3000.00 | 12000.00 | 85.77% | 0.00 | 33.50 | 185860.50 | 73.90 | 3.90 | 99.93% |
| 46 | 53 | 74 | 21 | 5.0 | 2.41 | 10000.00 | 4000.00 | 174796.00 | 3000.00 | 12000.00 | 85.77% | 7.30 | 38.80 | 183386.00 | 907.40 | 5.20 | 99.39% |
| 47 | 39 | 63 | 24 | 4.0 | 4.10 | 20000.00 | 2000.00 | 167381.00 | 5000.00 | 17000.00 | 79.18% | 0.00 | 34.60 | 181513.70 | 4015.70 | 8.40 | 97.62% |
| 48 | 68 | 89 | 21 | 5.0 | 3.15 | 20000.00 | 2000.00 | 167381.00 | 5000.00 | 17000.00 | 79.18% | 0.00 | 31.10 | 183078.20 | 189.80 | 7.00 | 99.84% |
| 49 | 59 | 87 | 28 | 4.0 | 4.81 | 30000.00 | 3000.00 | 166867.00 | 4000.00 | 8000.00 | 78.76% | 8.30 | 24.90 | 185436.20 | 667.30 | 6.20 | 99.51% |
| 50 | 51 | 85 | 34 | 1.0 | 3.67 | 30000.00 | 3000.00 | 166867.00 | 4000.00 | 8000.00 | 78.76% | 10.00 | 26.00 | 186356.70 | 578.40 | 7.00 | 99.57% |
| 51 | 63 | 82 | 19 | 2.0 | 7.45 | 32396.40 | 2180.00 | 144076.00 | 29697.00 | 22387.00 | 62.44% | 4.20 | 24.10 | 186215.00 | 3674.30 | 12.60 | 97.43% |
| 52 | 74 | 85 | 11 | 3.0 | 4.35 | 51271.06 | 7334.71 | 122249.24 | 46510.31 | 24657.02 | 48.51% | 6.10 | 14.20 | 179856.80 | 6361.60 | 6.10 | 96.00% |
| 53 | 58 | 89 | 31 | 2.0 | 3.31 | 120431.89 | 3023.09 | 98236.54 | 42933.13 | 13546.03 | 35.32% | 6.30 | 15.70 | 182497.00 | 8648.00 | 1.00 | 94.79% |
| 54 | 80 | 94 | 14 | 6.0 | 4.80 | 225715.36 | 6724.17 | 68341.01 | 1723.77 | 23876.27 | 20.94% | 0.00 | 14.50 | 185181.30 | 24.80 | 2.10 | 99.97% |
| 55 | 79 | 91 | 12 | 5.0 | 7.39 | 168037.45 | 2873.18 | 97690.33 | 7441.29 | 13459.27 | 33.74% | 0.00 | 0.00 | 183267.20 | 933.30 | 6.20 | 99.40% |
| 56 | 74 | 88 | 15 | 6.0 | 4.83 | 175192.84 | 6299.72 | 91160.26 | 10021.65 | 14518.42 | 30.67% | 0.00 | 0.00 | 181579.90 | 863.80 | 4.10 | 99.42% |
| 57 | 77 | 89 | 12 | 2.0 | 5.47 | 178078.04 | 7842.18 | 94144.23 | 8860.46 | 4958.26 | 32.03% | 0.00 | 0.00 | 182099.90 | 3071.70 | 3.10 | 98.08% |
| 58 | 73 | 90 | 17 | 6.0 | 9.35 | 183016.95 | 1209.36 | 90804.20 | 8188.29 | 13355.88 | 30.62% | 6.20 | 0.00 | 185393.60 | 1172.70 | 6.20 | 99.24% |
| 59 | 75 | 88 | 13 | 2.0 | 7.63 | 196513.85 | 4156.54 | 86924.08 | 6357.00 | 8021.25 | 28.79% | 4.20 | 0.00 | 191913.20 | 2428.90 | 3.10 | 98.51% |
| 60 | 12 | 70 | 58 | 2.0 | 3.23 | 76.30 | 0.00 | 173546.70 | 25.40 | 701.70 | 99.34% | 68.40 | 0.00 | 188664.70 | 12.40 | 0.00 | 99.96% |
| 61 | 15 | 57 | 42 | 3.0 | 2.66 | 33485.40 | 51.70 | 161440.20 | 1299.70 | 3612.50 | 80.74% | 1050.90 | 54.10 | 184417.20 | 154.60 | 54.10 | 99.29% |

| 62 | 70 | 88 | 18 | 3.0 | 3.02 | 33485.40 | 51.70 | 161440.20 | 1299.70 | 3612.50 | 80.74% | 0.00 | 56.20 | 174186.00 | 38.50 | 47.40 | 99.91% |
| --- | --- | --- | --- | --- | --- | --- | --- | --- | --- | --- | --- | --- | --- | --- | --- | --- | --- |
| 63 | 53 | 74 | 21 | 5.0 | 2.50 | 33485.40 | 51.70 | 161440.20 | 1299.70 | 3612.50 | 80.74% | 48.20 | 58.20 | 174048.80 | 60.20 | 50.20 | 99.88% |
| 64 | 20 | 63 | 43 | 2.0 | 2.31 | 46546.60 | 180.40 | 159220.40 | 2990.10 | 12565.90 | 71.79% | 107.20 | 28.90 | 167923.10 | 122.20 | 50.40 | 99.81% |
| 65 | 15 | 57 | 42 | 3.0 | 2.63 | 33485.40 | 51.70 | 161440.20 | 1299.70 | 3612.50 | 80.74% | 1477.50 | 34.30 | 177942.00 | 100.00 | 38.20 | 99.08% |
| 66 | 63 | 79 | 16 | 4.0 | 2.36 | 33485.40 | 51.70 | 161440.20 | 1299.70 | 3612.50 | 80.74% | 24.00 | 20.30 | 173825.50 | 34.20 | 12.90 | 99.94% |
| 67 | 25 | 70 | 45 | 4.0 | 1.83 | 16947.30 | 14.20 | 164018.30 | 6224.90 | 1107.20 | 86.96% | 4365.80 | 1.90 | 160674.10 | 7825.70 | 46.80 | 90.02% |
| 68 | 65 | 85 | 20 | 4.0 | 1.80 | 16947.30 | 14.20 | 164018.30 | 6224.90 | 1107.20 | 86.96% | 393.50 | 5.10 | 171037.60 | 464.30 | 18.20 | 99.34% |
| 69 | 43 | 70 | 27 | 3.0 | 9.15 | 88884.80 | 4018.83 | 140925.27 | 4400.08 | 2773.11 | 58.47% | 17.40 | 0.00 | 167881.20 | 1539.30 | 10.60 | 99.07% |
| 70 | 43 | 70 | 27 | 3.0 | 8.97 | 2035.27 | 3951.03 | 181681.60 | 4578.87 | 2533.65 | 93.28% | 14.90 | 0.00 | 168048.90 | 1614.10 | 4.60 | 99.03% |
| 71 | 54 | 72 | 18 | 5.0 | 4.22 | 148297.97 | 1879.29 | 114472.61 | 2631.54 | 3638.15 | 42.25% | 26.90 | 3.00 | 174514.60 | 1628.40 | 0.00 | 98.71% |
| 72 | 54 | 72 | 18 | 5.0 | 4.20 | 2082.56 | 1414.06 | 183178.89 | 2654.75 | 3686.15 | 94.90% | 36.50 | 40.40 | 178295.60 | 1023.60 | 42.40 | 99.16% |
| 74 | 18 | 73 | 55 | 2.0 | 9.55 | 16893.75 | 62.21 | 179054.30 | 241.30 | 1610.29 | 90.49% | 71.20 | 68.30 | 147452.00 | 111.50 | 614.40 | 99.38% |
| 75 | 51 | 84 | 33 | 2.0 | 2.28 | 0.00 | 0.00 | 185667.26 | 1949.15 | 2275.79 | 97.78% | 7.50 | 89.30 | 153976.00 | 1070.50 | 115.60 | 98.84% |
| 76 | 51 | 84 | 33 | 2.4 | 2.27 | 130085.38 | 0.00 | 124803.46 | 1820.96 | 2094.83 | 48.22% | 13.80 | 71.80 | 153333.00 | 1178.00 | 70.80 | 98.76% |
| 78 | 63 | 87 | 24 | 3.2 | 3.30 | 26709.30 | 92.90 | 142542.00 | 4027.80 | 1269.30 | 81.38% | 1226.00 | 104.00 | 164800.00 | 122.00 | 154.00 | 98.95% |
| 79 | 37 | 74 | 37 | 4.2 | 2.90 | 26709.30 | 92.90 | 142542.00 | 4027.80 | 1269.30 | 81.38% | 1870.00 | 111.00 | 158910.00 | 3600.00 | 83.0 | 95.66% |
| 80 | 29 | 58 | 29 | 1.7 | 3.60 | 26709.30 | 92.90 | 142542.00 | 4027.80 | 1269.30 | 81.38% | 1551.00 | 91.00 | 155480.00 | 2694.00 | 45.00 | 96.54% |

**Table S5:** Experimental parameters for Figure 5, detailing conditions of a validation experiment illustrating the impact of cold reactor temperature on Mg impurity reduction.

| Exp. # | T cold (°C) | T hot (°C) | ΔT (°C) | Flow rate (mL/min) | slurry concentration (g total solid per 100 mL) | Initial Ca (ppm) | Initial K (ppm) | Initial Li (ppm) | Initial Mg (ppm) | Initial Na (ppm) | Initial Li2CO3 purity (%) | Final Ca (ppm) | Final K (ppm) | Final Li (ppm) | Final Mg (ppm) | Final Na (ppm) | Final Li2CO3 purity (%) |
| --- | --- | --- | --- | --- | --- | --- | --- | --- | --- | --- | --- | --- | --- | --- | --- | --- | --- |
| 38 | 68 | 89 | 21 | 5.0 | 6.08 | 196365.87 | 764.40 | 84126.47 | 743.17 | 9650.55 | 28.84% | 12.70 | 40.10 | 185052.10 | 8.80 | 5.90 | 99.94% |
| 39 | 10 | 83 | 73 | 5.0 | 6.12 | 196365.87 | 764.40 | 84126.47 | 743.17 | 9650.55 | 28.84% | 132.10 | 34.60 | 186958.60 | 683.80 | 10.50 | 99.52% |

**Table S6:** Specifications of Machine Learning Models Used.

This study primarily utilized Gaussian Process (GP) models. The initial predictive model was a Gaussian Process Regressor (GPR), used to predict final impurity concentrations based on input parameters. For binary classification of outcomes (battery-grade vs. non-battery-grade), a standard Gaussian Process Classifier (GPC) was employed to define the decision boundary. Due to the GPC's kernel complexity, an alternative approach using a simpler GPR model for classification was also tested, which yielded comparable performance and offered greater flexibility via the alpha parameter. Finally, the NSGA-II algorithm was used for multi-objective optimization to identify the Pareto-optimal front for minimizing Mg and Ca concentrations.

| Task | Model | Specifications | Description |
|---|---|---|---|
| Regression | GPR (Scikit Learn) | Kernel : Matern(length_scale = 1, nu = 1.5)<br>alpha : 1e-10<br>n_restarts_optimizer : 10 | GPR is used for regression to predict final Mg and Ca concentrations, leveraging its ability to capture complex, non-linear relationships in the data. |
| Classification | GPC (Scikit Learn) | Kernel= ConstantKernel(1.0) * (Matern(length_scale = [0.3, 0.3], nu =1.5) + WhiteKernel(noise_level = 0.06))<br>n_restarts_optimizer : 10<br>max_iter_predict : 100 | GPC is used for binary classification, outputting probabilities and mimicking GPR's flexibility. |
| | GPR (Scikit Learn) | Kernel: RBF(length_scale=[0.3, 0.3])<br>Alpha: 0.06<br>Threshold: 0.5 | GPR classification uses thresholding of continuous output, leveraging simplicity and noise modeling. |
| Pareto frontier extraction | NSGA-II (Platypus-Opt) | Population Size: 500<br>Generations: 10000<br>Objectives: Minimize Mg and Ca concentrations | NSGA-II explores trade-offs between objectives, refining the Pareto front with diverse population. |

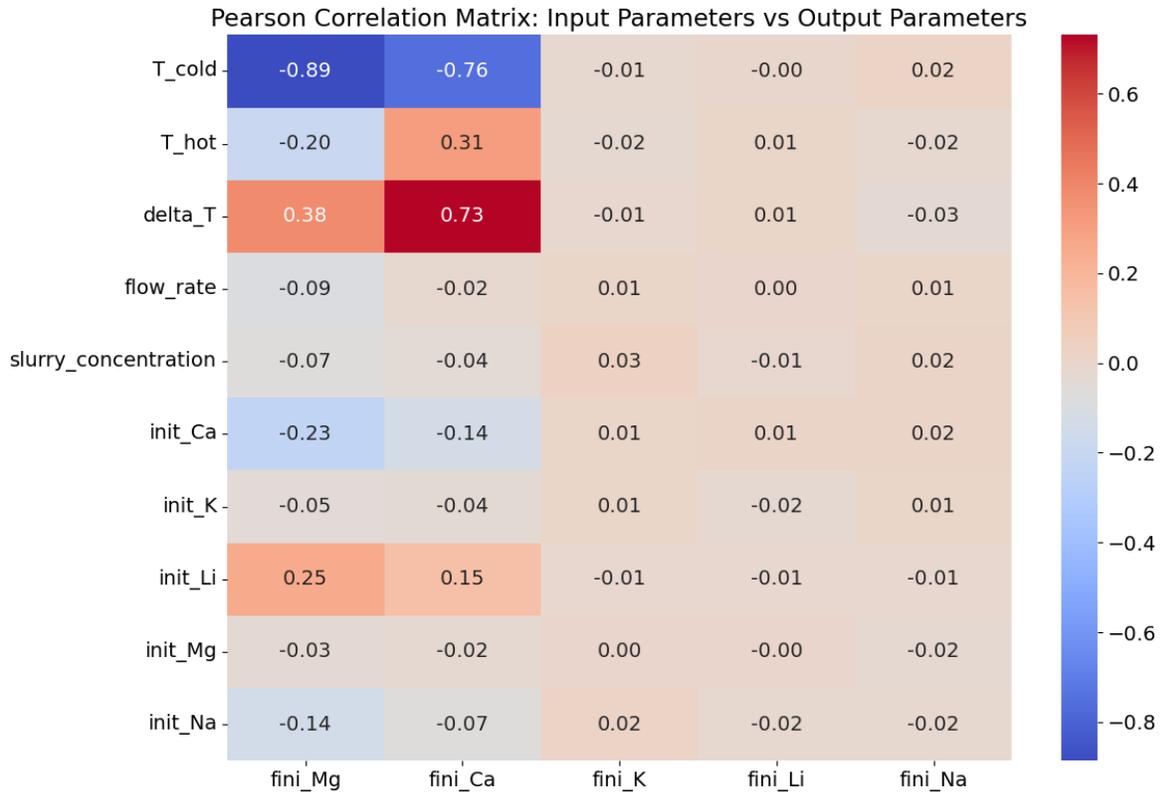

**Figure S3:** Pearson correlation matrix for GPR predictions on the 5000 generated surrogate space points using random walkers analyzing relationships among input parameters initial impurity concentrations and post-refinement impurity concentrations. Demonstrating strong predicted inverse correlation between temperature of the cold reactor and post refinement Mg concentration.

**References:**

1. China YS/T 582-2006 (Older Standard). Beijing China: Standards Press of China; 2006.

2. China YS/T 582-2013 (Current Standard) [Internet]. Beijing China: Standards Press of China; 2013 [cited 2025 Jun 15]. Available from: https://www.chinesestandard.net/Detail/237255/